\documentclass[preprint,12pt,authoryear]{elsarticle}
\usepackage[utf8]{inputenc}
\usepackage{booktabs}
\usepackage{caption}
\usepackage[table]{xcolor}
\usepackage[hidelinks]{hyperref}
\graphicspath{ {./images/} }
\usepackage{subcaption}

\usepackage{amssymb}
\usepackage{amsmath}

\usepackage[inline]{enumitem}

\usepackage{cleveref}

\usepackage[htt]{hyphenat}

\usepackage{siunitx}
\usepackage{xspace}

\newcommand{\terra}{\textsc{terra}\xspace}

\newcommand{\RQOne}{How do the count and fraction of flaky tests change between \terra releases?\xspace}
\newcommand{\RQTwo}{How persistent are individual flaky test cases across \terra releases?\xspace}
\newcommand{\RQThree}{What is the probability of detecting a flaky test given a specific number of executions?\xspace}
\newcommand{\RQFour}{How are failure intensities distributed across releases?\xspace}
\newcommand{\RQFive}{How are flaky tests distributed across \terra subcomponents?\xspace}
\newcommand{\RQSix}{How do flaky tests emerge, dissipate, or recur across the sequence of \terra releases?\xspace}

\newcommand{\COne}{Rarely Flaky\xspace}
\newcommand{\CTwo}{Persistently Flaky\xspace}
\newcommand{\CThree}{Intermittently Flaky\xspace}

\begin{document}

\begin{frontmatter}

\title{A Large-Scale Dynamic Characterization of Flaky Tests in Quantum Software: The Qiskit Terra Case Study}

\author[aff1]{Dongchan Kim\fnref{eq}}
\author[aff2]{Hamidreza Khoramrokh\fnref{eq}}
\author[aff1]{Lei Zhang}
\author[aff2]{Andriy Miranskyy}

\fntext[eq]{These authors contributed equally to this work.}

\affiliation[aff1]{organization={University of Maryland, Baltimore County},
            city={Baltimore},
            state={MD},
            country={USA}}

\affiliation[aff2]{organization={Toronto Metropolitan University},
            city={Toronto},
            state={ON},
            country={Canada}}

\begin{abstract}
Flaky tests, tests that pass or fail nondeterministically without changes to code or environment, pose a threat to software reliability. While classical software engineering has developed a body of techniques to study flakiness, evidence for quantum software remains limited. Prior work relies on static analysis or small sets of manually reported incidents, leaving open questions about their prevalence, characteristics, and detectability.

This paper presents a large-scale dynamic characterization of flaky tests in quantum software through a longitudinal case study of the Qiskit \terra core library. We executed the Qiskit \terra test suite 10,000 times across 23 releases in controlled environments. For each release, we measured test-outcome variability, identified flaky tests, estimated empirical failure probabilities, analyzed recurrence across versions, used Wilson confidence intervals to quantify baseline rerun budgets, and mapped flaky tests to \terra subcomponents.

Across 27,026 qualified test identifiers, we identified 62 unique flaky tests. Although flakiness rates were low (0--0.17\%), recurrence was substantial: 52 of 62 flaky tests (83.87\%) reappeared in multiple releases, while only 10 tests (16.13\%) were confined to a single release. Failure probabilities spanned several orders of magnitude, with a median of $9 \times 10^{-4}$ and 34 tests (54.84\%) at or below $10^{-3}$, implying that thousands to tens of thousands of executions may be required for confident detection.

These results show that flakiness observed in the studied \terra releases is rare under our controlled protocol but difficult to detect under typical continuous integration budgets. To support future research and replication, we release a dataset of per-test execution outcomes.
\end{abstract}

\begin{keyword}

quantum flaky tests \sep dynamic testing \sep quantum software testing \sep quantum software engineering
\end{keyword}

\end{frontmatter}

\section{Introduction}\label{sec:intro}

Flaky tests, i.e., tests that unpredictably alternate between passing and failing despite no changes to code or environment, pose a persistent threat to software reliability~\citep{luo2014empirical, micco2017state, memon2017taming}. Their presence reduces trust in test outcomes, obscures real defects, and slows continuous integration (CI) or continuous deployment (CD) pipelines. In classical software, extensive research has shown that flakiness arises from diverse sources such as concurrency, nondeterministic execution order, platform dependencies, and environment-related timing issues~\citep{gruber2022survey,parry2021survey}. To mitigate these challenges, the classical literature has developed a spectrum of techniques: dynamic re-execution strategies~\citep{alshammari2021flakeflagger,lam2019idflakies,bell2018deflaker}, static analysis~\citep{luo2014empirical, gruber2021empirical, lam2020study, lam2019root, parry2021survey}, and machine-learning-based predictors~\citep{ziftci2020flake, bell2018deflaker, lam2019idflakies, dutta2020detecting, alshammari2021flakeflagger, verdecchia2021know}.

As the quantum computing ecosystem matures, similar reliability challenges have begun to surface~\citep{murillo2025quantum}. Frameworks such as Qiskit~\citep{qiskit2024} now provide full software stacks and have testing infrastructures comparable in scale to classical systems. Quantum software combines familiar sources of software nondeterminism with domain-specific algorithms and abstractions, including probabilistic program semantics, noise-aware compilation flows, floating-point approximations, evolving toolchains, and constraint-sensitive transpiler optimizations. These mechanisms are not necessarily unique to quantum software, but they arise in implementations of quantum circuits, operators, synthesis, and compilation. Early studies have identified flaky tests in quantum software repositories~\citep{zhang2023identifying}, but current work relies primarily on static analysis of small, manually curated, incident-report-driven datasets~\citep{zhang2024automated, kaur2025identifying}.
 Thus, we still lack a systematic, empirical understanding of how frequently flaky tests occur in quantum software, how they evolve across releases, and how difficult they are to detect given realistic rerun budgets.

Static machine-learning approaches depend critically on the availability of reliable labeled data. Without large, empirically derived datasets, especially those capturing very low-probability events, such models risk overfitting to limited historical cases~\citep{verdecchia2021know}. Moreover, the very small failure probabilities observed in some quantum-software tests may make the limited rerun budgets typically used in CI pipelines insufficient to reveal rare nondeterministic behavior.

To close this gap, we conduct a large-scale dynamic case study of flaky tests in quantum software. We repeatedly execute the complete test suite for the Qiskit \terra component\footnote{\label{note:terra} From hereon we will refer to it as \terra. Its role is summarized in the project documentation: ``This library is the core component of Qiskit, which contains the building blocks for creating and working with quantum circuits, quantum operators, and primitive functions (Sampler and Estimator). It also contains a transpiler that supports optimizing quantum circuits, and a quantum information toolbox for creating advanced operators.''~\citep{qiskit_description}} 10,000 times across 23 releases in controlled, containerized environments. Our study adopts a high-budget repeated-execution protocol established in classical flaky-test research and applies it longitudinally to a quantum-software framework. This design surfaces rare nondeterministic failures that would remain undetected under conventional testing strategies and enables empirical analysis of failure probabilities, temporal persistence, and subcomponent-level patterns. We also validate our initial findings and experimental settings with Qiskit developers~\citep{qiskit2026validation_discussion}.

We formally state the \textit{research questions} (RQs) guiding this study.
\begin{enumerate}[leftmargin=*, label=\textbf{RQ\arabic*:}]
\item \RQOne
\item \RQTwo
\item \RQThree
\item \RQFour
\item \RQFive
\item \RQSix
\end{enumerate}
This work makes the following \textit{contributions}.
\begin{enumerate}
    \item \textit {Large-scale empirical analysis.} We rerun the \terra test suite 10,000 times across 23 \terra releases (approximately 70 CPU-years of computation), providing execution-level evidence about how observed flakiness changes and recurs over time.

    \item \textit{Empirical calibration of detectability.} We quantify empirical failure probabilities for the flaky tests observed in \terra and apply standard binomial and Wilson calculations to estimate baseline rerun requirements under an independent and identically distributed (IID) model. This analysis translates the measured failure-probability distribution into practical implications for CI rerun budgets; it is not presented as a new statistical method.

    \item \textit{Public execution-level dataset.} We release a curated dataset containing 62 unique flaky tests observed in \terra, annotated with results from 10,000 controlled executions per release. The dataset provides a dynamic benchmark under the specified protocol for future research in quantum test reliability, debugging, replication, and machine-learning prediction. It is available at \url{https://zenodo.org/records/20500432}.

\end{enumerate}

By integrating large-scale dynamic evidence with longitudinal and probabilistic analysis, this case study characterizes flaky-test behavior in a major quantum-software library. The findings highlight the practical challenges in detecting low-probability nondeterminism and provide a foundation for replication and tool evaluation in quantum-software testing.

The remainder of the paper is organized as follows. \Cref{sec:related} reviews related work. \Cref{sec:design} describes the software under study, data-collection process, and execution environment. \Cref{sec:results} presents the empirical analysis and answers the RQs. \Cref{sec:discussion} synthesizes the findings and discusses their practical implications. \Cref{sec:threats} discusses threats to validity. \Cref{sec:conclusions} concludes and outlines future work.

\section{Related Work}\label{sec:related}

Flaky tests have been widely studied in classical software and only recently examined in quantum software.
This section reviews both bodies of work, highlighting the methodological gap that motivates our dynamic study.

\subsection{Flaky Tests in Classical Software}
Research on classical flaky tests spans three main areas: dynamic detection, static prediction, and root-cause analysis as discussed below.

Early work on classical flaky tests focused on detecting them by rerunning tests under varying conditions or orders. Techniques such as iDFlakies~\citep{lam2019idflakies} repeatedly reorder and execute test suites to surface order-dependent flaky behavior, resulting in a dataset of 422 flaky tests. DeFlaker~\citep{bell2018deflaker} detects flaky failures in real time and flags failures not associated with recent code changes. Such methods remain effective but are computationally expensive, especially for large test suites.

To reduce rerun cost, several tools predict flakiness from code characteristics. For instance, \citet{pinto2020what} present a purely static approach that warns developers about flaky tests as they write code. FLAST~\citep{verdecchia2021know} represents test code in a high-dimensional space using bag-of-words features and applies sparse random projection with $k$-nearest neighbors. FlakeFlagger~\citep{alshammari2021flakeflagger} combines static and dynamic features, supported by a dataset created by running 811 tests 10,000 times~---~far larger than previous dynamic efforts. They show that small rerun budgets (10, 100, and 1000 re-runs) detect only a minority of flaky tests ($\approx$ 26\%, 45\%, and 67\%, respectively), motivating higher-budget studies.

\citet{luo2014empirical} provide a widely used taxonomy of ten flaky-test root causes, with asynchronous waits, concurrency, and order dependencies accounting for most cases. \citet{eck2019understanding} discovered four new causes of flakiness, including those related to test environments and unmet assumptions that had not been reported previously.

\subsection{Flaky Tests in Quantum Software}

Flaky tests in quantum software have received limited attention, with existing research relying almost exclusively on static or manual methods.

\citet{zhang2023identifying, zhang2024automated} performed the first empirical study of flaky tests in quantum repositories by mining issue trackers and pull requests across 14 open-source projects. They identified 46 instances of flaky tests in 12 projects (0.3–1.8\% of reported bugs), categorized eight common causes of quantum test flakiness, and documented seven common fix strategies. This work demonstrates that flakiness already exists in quantum ecosystems, but the scale is restricted by the availability of human-reported incidents.

\citet{kaur2025identifying} proposed the first machine-learning-based flaky-test detector for quantum software. Using a bag-of-words representation similar to classical static predictors~\citep{verdecchia2021know}, they trained various classifiers that achieved strong performance on the limited set of available labeled cases. However, like classical static approaches, the accuracy of such models is fundamentally constrained by the lack of large, dynamic ground-truth datasets.

\citet{sivaloganathan2024automating,sivaloganathan2026automating} proposed an automated framework for detecting flaky tests in quantum software by mining issue reports and pull requests using embedding-based similarity and large language models, expanding prior manually curated datasets. While their approach achieves strong performance for flakiness detection from textual artifacts, it remains dependent on developer-reported incidents rather than direct observation of execution-level nondeterminism.

\subsection{Comparison with Prior Work}

\begin{table}[t]
\centering
\caption{Comparison of representative flaky-test studies. The table contrasts prior work by domain, evidence source, detection method, scale, and rerun budget. N/A indicates that the study did not use a dynamic test-reexecution protocol.}
\label{tab:related_work_comparison}
\scriptsize
\resizebox{\textwidth}{!}{%
\begin{tabular}{p{0.18\textwidth} p{0.10\textwidth} p{0.20\textwidth} p{0.19\textwidth} p{0.20\textwidth} p{0.17\textwidth}}
\toprule
Study & Domain & Evidence Source & Method & Scale & Rerun Budget \\
\midrule
\citet{luo2014empirical}
& Classical
& Commit logs, bug reports, and patches
& Manual/static root-cause analysis
& 201 flaky-test-fixing commits from 51 projects
& N/A \\
\midrule
\citet{lam2019idflakies} (iDflakies)
& Classical
& Test executions under different orders
& Dynamic detection via test reordering
& 422 flaky tests from 82 projects
& Variable; default 20 rounds; up to 56h/project \\
\midrule
\citet{verdecchia2021know} (FLAST)
& Classical
& Test source code and existing labeled flaky/non-flaky test datasets
& Static prediction using vector-space modeling, dimensionality reduction, and k-nearest neighbors
& 1,383 flaky tests and 26,702 non-flaky tests from 13 projects
& N/A \\
\midrule
\citet{alshammari2021flakeflagger}  (FlakeFlagger)
& Classical
& Repeated test-suite executions
& Dynamic dataset construction + ML-based prediction
& 811 flaky tests from 24 projects
& 10,000 reruns/project \\
\midrule
\citet{zhang2023identifying,zhang2024automated}
& Quantum
& Issue reports and pull requests
& Issue/PR mining and manual analysis
& 46 flaky-test reports from 12 of 14 projects
& N/A \\
\midrule
\citet{kaur2025identifying}
& Quantum
& Existing labeled quantum flaky-test cases
& Static ML prediction
& 45 flaky Python files and 243 non-flaky Python files
& N/A \\
\midrule
\citet{sivaloganathan2024automating,sivaloganathan2026automating}
& Quantum
& Issue reports and pull requests
& Embedding-based mining + LLM classification
& 71 flaky-test reports from 12 of 14 projects
& N/A \\
\midrule
\textbf{This work}
& \textbf{Quantum}
& \textbf{Repeated test-suite executions}
& \textbf{Dynamic characterization}
& \textbf{1 project / 23 releases; 348 occurrences / 62 unique tests}
& \textbf{10,000 runs/release} \\
\bottomrule
\end{tabular}
}
\end{table}

Table~\ref{tab:related_work_comparison} positions our study relative to representative flaky-test studies in both classical and quantum software. The comparison is not intended as a direct numerical ranking, because prior studies differ in their units of analysis and data sources: static machine-learning studies depend on pre-existing labeled cases, whereas dynamic studies identify flakiness through repeated test execution. Instead, the table highlights the empirical gap addressed by this work. Prior quantum studies have shown that flaky tests exist in quantum software, but they do not provide large-scale execution-level observations, empirical failure probabilities, temporal recurrence patterns, or rerun-budget calibration. Our study complements these efforts by applying an established high-budget rerun design across 23 Qiskit \terra releases.

\subsection{Summary and Gap}

Across both domains, dynamic rerun studies have proven essential for uncovering rare nondeterministic failures and for building execution-level benchmark datasets. Yet, no such large-scale dynamic analysis exists for quantum software. Prior quantum studies rely on incident reports or static features and therefore cannot:
\begin{enumerate*}[label=(\roman*)]
    \item measure flakiness observed through repeated execution across versions,
    \item quantify empirical failure probabilities, or
    \item assess detectability under realistic rerun budgets.
\end{enumerate*}

This work addresses that gap through a large-scale dynamic case study in a quantum-software framework. The contribution lies in the scale, longitudinal evidence, and public dataset rather than a new detection procedure. The dataset provides an empirical foundation for future statistical analysis, replication, and ML-based prediction.

\section{Study Design}\label{sec:design}
This study aims to measure and characterize flaky behavior in a quantum-software case through large-scale dynamic re-execution. Our study design was adopted from the high-budget dynamic rerun protocol used in FlakeFlagger~\citep{alshammari2021flakeflagger}, which demonstrated the value of repeated execution for constructing flaky-test benchmark data. We adapt this established experimental style to 23 Qiskit \terra releases, using controlled containerized environments and analyzing release-level failure probabilities, recurrence, and subcomponent-level patterns. We do not propose a new flaky-test detector or statistical method.

This section describes the software under study, our data-collection pipeline, and the execution environment used to generate the dataset.

\subsection{Software Under Study}\label{subsec:sus}

We focus on the Qiskit software stack~\citep{qiskit2024}, one of the most widely used frameworks for quantum circuit construction, compilation, and analysis~\citep{khan2025mining}. Qiskit provides high-level circuit APIs, a multi-stage transpiler, simulators, backend interfaces, and a comprehensive built-in test suite, making it an important and mature case-study target for nondeterministic behavior in quantum software. We do not treat it as statistically representative of the broader quantum-software ecosystem.

Our analysis concentrates on the core Qiskit library~\citep{qiskit2024}, formerly known as \texttt{qiskit-terra} prior to v.1.0, renamed \texttt{qiskit} thereafter. Its role is summarized in \Cref{note:terra}.

\terra primarily exercises circuit construction, quantum-information routines, synthesis, and compilation/transpilation on classical computing infrastructure. Frameworks with comparable compiler and simulator layers may share concerns such as stochastic optimization, seed propagation, numerical tolerance, and topology-constrained routing. Their detailed behavior may nevertheless differ with the programming model, implementation language, compiler architecture, target gate set, and hardware topology. Other parts of the quantum-software ecosystem (especially hardware-backed runtimes and cloud-service clients) may additionally be affected by network variability, queueing, calibration drift, service availability, and device noise, none of which is characterized by this study. Accordingly, the experimental protocol can be replicated elsewhere, but the measured rates and root-cause distribution require independent replication before being generalized beyond \terra.

We study all 23 releases between v.0.25.0 and v.1.2.4 (see \Cref{tab:count_fraction}). The lower bound reflects an engineering constraint (the earliest versions rely on outdated toolchains that no longer build reproducibly) while the upper bound coincides with the last release supporting Python~3.8, which ensures compatibility across all versions under study.

\subsection{Data Collection Method and Execution Testbed}\label{subsec:data}

Our goal is to obtain high-fidelity, release-specific measurements of nondeterministic test outcomes. To achieve this, we designed a pipeline that repeatedly executes the \terra test suite 10,000 times per release in clean, isolated environments.

Our execution protocol is designed to measure flakiness under a controlled, repeatable rerun setting rather than to maximize flake discovery through perturbation. We do not reorder tests, inject timing delays, vary resource stress deliberately, or otherwise perturb the execution environment. Consequently, the observed counts and rates characterize flakiness detected under this controlled protocol, not all possible flakiness in \terra. They should be interpreted as conservative lower bounds on behavior that might be exposed across a broader range of orders, timing conditions, stress levels, platforms, dependency states, and hardware interactions. Increasing the repetition count improves sensitivity to rare events within the sampled conditions, but it cannot recover mechanisms that the protocol does not activate.

Qiskit uses TOX software~\citep{tox} to virtualize the test environment and execute test cases on specific platforms. Since Linux is both Qiskit's primary development platform and the operating system of the high-performance computing (HPC) cluster that we used to run test cases, we selected it for our study.

Each run of the test suite was executed in a clean and isolated environment to ensure that the environment's state did not affect the tests. For each version, we prepared a dedicated Dockerfile that pins Python~3.8 and installs all dependencies needed to build and run that version’s test suite. Because Qiskit depends on the Rust toolchain~\citep{matsakis2014rust}, we selected a compatible Rust toolchain for each release based on Qiskit's evolving minimum requirements. Although required libraries were cached in the Docker images, internet access was still necessary because tox, Qiskit’s test orchestrator, resolves and verifies dependencies at runtime before executing the test suites.

Our HPC environment supports Singularity~\citep{kurtzer2017singularity} rather than Docker containers. Accordingly, each Docker image was converted to a Singularity image using the Singularity CLI. We then scheduled 11,000 executions per image, each using 4~CPU cores (Intel Broadwell processors) and 24~GB of RAM. Gathering the data on our testbed required approximately 70 CPU-years.

The slight over-provisioning (11,000 runs) provides resilience against transient issues such as network failures or dependency-resolution timeouts. After execution, we retained only the first 10,000 successful runs per release to maintain strict comparability across versions.
This retained budget balances detection power and computational cost. It is consistent with prior high-budget dynamic flaky-test studies, such as~\citet{alshammari2021flakeflagger}, while substantially exceeding the rerun counts typically feasible in per-commit CI.\footnote{
Smaller budgets would reduce cost but would make rare flaky failures much less likely to appear in the retained data. Larger budgets would improve detection but with rapidly increasing computational cost. We therefore selected $10{,}000$ retained executions per release as a practical compromise: high enough to expose failures at the $10^{-4}$ empirical level when they occur in the sample, yet still tractable for a 23-release study that already required about 70 CPU-years.

We quantify this trade-off in \Cref{sec:theoretical_true_p}. There, we derive the detection probability and use Wilson lower confidence bounds to estimate conservative rerun budgets. The analysis shows that $5{,}000$ executions would often miss failures around the $10^{-4}$ level, $20{,}000$ executions would still not provide confident detection, and confidence-based budgets can reach the order of $10^5$ executions. Thus, $10{,}000$ executions should be read as a high but tractable dynamic sampling budget, not as a guarantee of detecting all rare flaky tests.
}

After the initial high-volume full-suite executions, we also performed targeted reruns as an internal validation step. We aggregated a master list of candidate flaky tests, including tests that required rechecking because of possible infrastructure- or parallelization-related behavior. For each release, we then selectively reran only the candidate tests that were present in that version. This procedure was part of our experimental workflow rather than a proposed CI framework, but it illustrates a practical strategy for reducing unnecessary computation: broad discovery runs can identify candidate flaky tests, while subsequent validation runs can prioritize those candidates instead of rerunning the entire test suite at the same depth.

Each test suite produces text-formatted output in the standard \texttt{pytest} style. We collected all outputs into version-specific folders and parsed them using custom Python scripts. These scripts extract per-test pass/fail/skip counts and consolidate them into a structured CSV file. This process yields one CSV table per release, each summarizing approximately 20,000 test cases.

\subsection{Dataset Description}\label{subsec:env}

All collected artifacts are publicly available at \url{https://zenodo.org/records/20500432}. Each of the 23 CSV files (one per release) shares a common schema summarized in \Cref{tbl:schema}. The fields describe per-test execution outcomes and the environment used for that release.

\begin{table}[ht!]
\centering
\caption{Schema of the per-release execution dataset. Each row corresponds to one test case in one \terra release.}
\begin{tabular}{p{0.28\linewidth} p{0.65\linewidth}}
\toprule
Attribute & Description \\
\midrule
\texttt{test\_case\_name} & The identifier of the individual \terra test case. \\
\texttt{passed\_attempts} & Number of runs in which the test passed. \\
\texttt{failed\_attempts} & Number of runs in which the test failed. \\
\texttt{skipped\_runs} & Number of skipped runs (e.g., via skip markers or environment constraints); skips were outside our control. \\
\texttt{total\_runs} & The total number of executions attempted for the test case; computed as
$\texttt{passed\_attempts} + \texttt{failed\_attempts} + \texttt{skipped\_runs}$; equals 10,000 in all retained rows. \\
\texttt{success\_rate} & The fraction of successful executions; computed as
$\texttt{passed\_attempts} / \texttt{total\_runs}$. \\
\texttt{qiskit\_version} & The \terra release tag used for the run.\\
\texttt{python\_version} & Python version used (always 3.8). \\
\texttt{rust\_version} & Rust toolchain version used to satisfy \terra build requirements for that release. \\
\bottomrule
\end{tabular}

\label{tbl:schema}
\end{table}
A test case is identified by its fully qualified \texttt{test\_case\_name}, consisting of the module path, class name, and method name. The 27{,}026 distinct test cases reported in \Cref{tab:count_fraction} correspond to the union of these fully qualified identifiers across all 23 releases, not to per-version test instances. Thus, a test appearing in multiple releases is counted once in the ``Unique'' row if it has the same fully qualified identifier. For cross-release recurrence analysis, we treat tests sharing the same fully qualified name as the same test. This is an identifier-level definition of uniqueness, not necessarily byte-level or semantic equivalence of the test body across releases. The underlying test body may evolve through code edits, so a ``persistent'' flaky test could in principle correspond to a sequence of slightly different implementations.

Our conceptual definition follows classical flaky-test research: a flaky test can produce inconsistent pass/fail outcomes for the same version of the code without a relevant code change. We operationalize this definition within each \terra release by labeling a test as \textit{flaky} when the controlled rerun sample contains at least one pass and at least one failure: formally, $\texttt{passed\_attempts} > 0$ and $\texttt{failed\_attempts} > 0$. The budget of 10,000 executions is not part of the definition; it determines the sensitivity of the observation protocol.

Classical flaky-test studies span multiple evidence-generation protocols rather than relying on a single standard procedure. Most directly comparable to our study, FlakeFlagger constructed its flaky-test oracle by executing each project test suite 10,000 times and identifying tests whose outcomes changed across executions \citep{alshammari2021flakeflagger}. iDFlakies also uses repeated execution, but additionally preserves, randomizes, or reverses test orders across configurations to expose and partially classify order-dependent and non-order-dependent behavior \citep{lam2019idflakies}. DeFlaker combines observed failures with coverage and code-change information \citep{bell2018deflaker}, whereas incident-mining studies derive evidence from developer reports and subsequent fixes.

Each protocol defines flakiness under a different set of observation conditions, so they surface overlapping but not identical sets of flaky tests. Ours is closest to the high-budget repeated execution used to build the FlakeFlagger dataset, but we apply it across 23 consecutive releases of a quantum software system, holding the test order fixed and the environment controlled. What we capture is therefore the outcome variability that surfaces under those sampled conditions. Order-dependent failures are a separate matter: exposing them requires running the suite in alternative orders, which we do not do. For each version, we count as flaky the test cases that meet this criterion.

\subsection{Validation with Qiskit Developers}\label{sec:validation}

To validate our experimental procedure, we shared our experimental setup and preliminary findings with IBM Qiskit developers and obtained practitioner feedback on our execution protocol and the interpretation of the detected flaky tests~\citep{qiskit2026validation_discussion}. The developers pointed out that our earlier configuration used a higher degree of parallelism than is typical in Qiskit's production CI/CD pipelines. This feedback led us to revise the experimental protocol so that the final reported results better reflect realistic Qiskit testing practice rather than an artificially aggressive stress-testing scenario. After reducing the level of parallelism and rerunning the experiments under the revised configuration, the observed level of flakiness decreased. Therefore, the results reported in this paper should be interpreted as flakiness detected under a controlled and practitioner-informed execution protocol rather than as an exhaustive estimate of all possible flaky behavior that could be exposed under more adversarial or highly parallel settings.

This interaction strengthened this study in two ways. First, it provides external validation from developers directly involved in the Qiskit ecosystem. Second, it clarifies the practical trade-off between maximizing flaky-test discovery and measuring flakiness under realistic operational conditions. More aggressive configurations may expose additional nondeterministic behavior, but they may also overestimate the flakiness experienced by developers in regular CI/CD workflows.

As an additional validation exercise, we applied the revised protocol to a more recent Qiskit version in a newer Python environment and identified an additional flaky test that was fixed by Qiskit developers~\citep{qiskit2026fix}. Although this recent-version result is not included in the main longitudinal analysis, which focuses on the 23 historical releases studied under a consistent Python 3.8 environment, it provides further evidence that dynamic re-execution remains applicable to the current Qiskit ecosystem.

\section{Study Results}\label{sec:results}

This section presents the empirical findings of our study. We analyze flakiness across the \terra releases, address each research question in turn, and then summarize report-level root-cause and fix-pattern observations.

\Cref{tab:count_fraction} summarizes the flakiness count, flakiness rate, and the relative contribution of each release to all observed flaky-test occurrences. Across the 23 releases, we observed 348 flaky-test occurrences, corresponding to 62 unique flaky tests out of 27,026 distinct test cases.

Let us now focus on individual research questions.

\begin{table}[ht!]
\centering
\caption{Test suite size and flakiness across \terra releases. For each release, we report the total number of test cases, the number of non-skipped tests, the number of flaky tests (i.e., those that both pass and fail across 10{,}000 runs),
the flakiness rate, and the contribution of that release to the global set of unique flaky tests.
The flakiness rate is computed as the ratio of flaky test count over total test count.
The share of flaky-test occurrences is computed as the ratio of the flaky-test count for a given release over the total number of flaky-test occurrences across all 23 releases.
The final two rows aggregate totals across all releases. }

\captionsetup{labelfont=bf} %
\rowcolors{2}{gray!10}{white} %
\resizebox{\columnwidth}{!}{%
\begin{tabular}{
  >{\raggedright\arraybackslash}p{2cm}
  >{\raggedleft\arraybackslash}p{2.5cm}
  >{\raggedleft\arraybackslash}p{2.5cm}
  >{\raggedleft\arraybackslash}p{2cm}
  >{\raggedleft\arraybackslash}p{2cm}
  >{\raggedleft\arraybackslash}p{2.5cm}
  >{\raggedleft\arraybackslash}p{2.5cm}
}
\toprule
Version & Total Test Count & Non-Skipped Tests Count & Flaky Tests Count & Flakiness rate (\%) & Share of Flaky-Test Occurrences (\%)\\
\midrule
0.25.0 & \num{21789} & \num{21445} & 0 & 0.00 & 0.00\\
0.25.1 & \num{21796} & \num{21303} & 12 & 0.06 & 3.45 \\
0.25.2 & \num{21813} & \num{21320} & 13 & 0.06 & 3.74 \\
0.25.3 & \num{21834} & \num{21339} & 12 & 0.06 & 3.45 \\
0.45.0 & \num{21931} & \num{21447} & 6 & 0.03 & 1.72 \\
0.45.1 & \num{21938} & \num{21455} & 21 & 0.10 & 6.03 \\
0.45.2 & \num{21947} & \num{21464} & 0 & 0.00 & 0.00 \\
0.45.3 & \num{21947} & \num{21464} & 18 & 0.08 & 5.17 \\
0.46.0 & \num{21975} & \num{21491} & 23 & 0.11 & 6.61 \\
0.46.1 & \num{21979} & \num{21494} & 25 & 0.12 & 7.18 \\
0.46.2 & \num{21989} & \num{21504} & 22 & 0.10 & 6.32 \\
0.46.3 & \num{21994} & \num{21509} & 15 & 0.07 & 4.31 \\
1.0.0 & \num{16316} & \num{16122} & 25 & 0.16 & 7.18 \\
1.0.1 & \num{16320} & \num{16126} & 27 & 0.17 & 7.76 \\
1.0.2 & \num{16333} & \num{16139} & 27 & 0.17 & 7.76 \\
1.1.0 & \num{17406} & \num{17208} & 18 & 0.10 & 5.17 \\
1.1.1 & \num{17418} & \num{17220} & 17 & 0.10 & 4.89 \\
1.1.2 & \num{17428} & \num{17229} & 16 & 0.09 & 4.60 \\
1.2.0 & \num{17668} & \num{17468} & 14 & 0.08 & 4.02 \\
1.2.1 & \num{17678} & \num{17474} & 21 & 0.12 & 6.03 \\
1.2.2 & \num{17684} & \num{17480} & 16 & 0.09 & 4.60 \\
1.2.3 & \num{17684} & \num{17480} & 0 & 0.00 & 0.00 \\
1.2.4 & \num{17684} & \num{17480} & 0 & 0.00 & 0.00 \\
\midrule
Total & \num{452551} & \num{444661} & 348 & 0.08 & 100.0 \\
Unique & \num{27026} & \num{26639} & 62 & & \\
\bottomrule
\end{tabular}
}
\label{tab:count_fraction}
\end{table}

\subsection{RQ1: \RQOne}

Table~\ref{tab:count_fraction} summarizes the number of flaky tests per release across the 23 \terra versions studied. We observe substantial variation: the number of flaky tests ranges from 0 (in v.0.25.0, v.0.45.2, v.1.2.3, and v.1.2.4) to 27 (in v.1.0.1 and v.1.0.2), with a mean of 15.1 (SD = 8.6) and a median of 16. Flakiness rates span 0--0.17\%, with a mean of 0.08\% (SD = 0.05\%).

Flakiness is not monotonic over time and shows episodic patterns. The highest absolute counts occur in mid-to-late releases, peaking in the v.0.46.x series (15--25 flaky tests) and the v.1.0.x series (25--27), rather than in the early v.0.25.x releases (0--13). Major-version transitions do not consistently coincide with increases in flakiness: the transition from v.0.46.3 (15) to v.1.0.0 (25) shows a clear jump, but the transition from v.0.25.3 (12) to v.0.45.0 (6) is associated with a decrease, and v.1.0.2 (27) to v.1.1.0 (18) again declines. Several releases interrupt otherwise active periods with zero observed flaky tests (v.0.45.2 between v.0.45.1 and v.0.45.3; v.0.25.0 at the start of its minor series), suggesting that flakiness can briefly disappear even within a stable architectural window.

The complete absence of observed flakiness in the final two releases (v.1.2.3 and v.1.2.4), preceded by a sequence of 14--21 flaky tests in v.1.1.x and v.1.2.0--v.1.2.2, may reflect a more sustained stabilization rather than a transient dip. Overall, the rate observed in the studied \terra releases under our protocol is episodic: it rises during periods of active development (notably the v.0.46.x and v.1.0.x series), drops sharply in occasional releases, and appears to stabilize in the most recent releases of our study window. As discussed in \Cref{sec:theoretical_true_p}, however, zero observed failures at a finite budget do not prove that flakiness has been eliminated.

\subsection{RQ2: \RQTwo}

The answer to the question is given in \Cref{tab:test-per-version}. We identified 62 unique flaky tests across the 23 releases. Our dataset shows a markedly different pattern: only 10 tests (16.13\%) appear in exactly one release, while the remaining 52 tests (83.87\%) recur across multiple releases.

The distribution is concentrated in the mid-range of persistence. The most frequent recurrence span is $k=5$ releases, accounting for 12 tests (19.35\%), followed by $k=7$ with 11 tests (17.74\%) and $k=6$ with 6 tests (9.68\%). Together, these three spans account for 29 tests (46.77\%) of all flaky tests, suggesting that a moderate but sustained level of recurrence is the norm rather than the exception in this dataset. Tests with very long persistence ($k \ge 10$) are rare, comprising only 5 tests (8.06\%), indicating that indefinite recurrence is uncommon. We will explore patterns of reappearance in \Cref{sec:individual_test_case_behaviour}.

\begin{table}[tbh]
\centering
\caption{Persistence of flaky tests across \terra releases. Each row shows the number of tests that were flaky in exactly $k$ \terra releases within our study window (v.0.25.0--v.1.2.4). The maximum observed persistence was $k=14$, i.e., no test was flaky in more than $14$ releases.}

\begin{tabular}{rrr}
\toprule
Number of Releases ($k$) & Flaky Tests Count & Flaky Tests (\%)   \\
\midrule
1 & 10 & 16.13 \\
2 & 3 & 4.84\\
3 & 2 & 3.23 \\
4 & 4 & 6.45 \\
5 & 12 & 19.35 \\
6 & 6 & 9.68 \\
7 & 11 & 17.74 \\
8 & 6 & 9.68 \\
9 & 3 & 4.84 \\
10 & 1 & 1.61 \\
12 & 1 & 1.61 \\
13 & 2 & 3.23 \\
14 & 1 & 1.61 \\

\midrule
Total & 62 & 100.00 \\
\bottomrule
\end{tabular}
\label{tab:test-per-version}
\end{table}

\subsection{RQ3: \RQThree}
Here, we answer RQ3 using two approaches: an empirical and a theoretical approach.

\subsubsection{Empirical analysis}
For each test case, we compute the empirical failure probability
\begin{equation*}
  \hat p_i = \frac{F_i}{N}, \qquad i=1,\ldots,62,
\end{equation*}
where $F_i$ is the number of failed runs of test $i$ and $N=10000$ is the total number of runs of test $i$.

\begin{figure}[tbh]
    \centering
    \includegraphics[width=\textwidth]{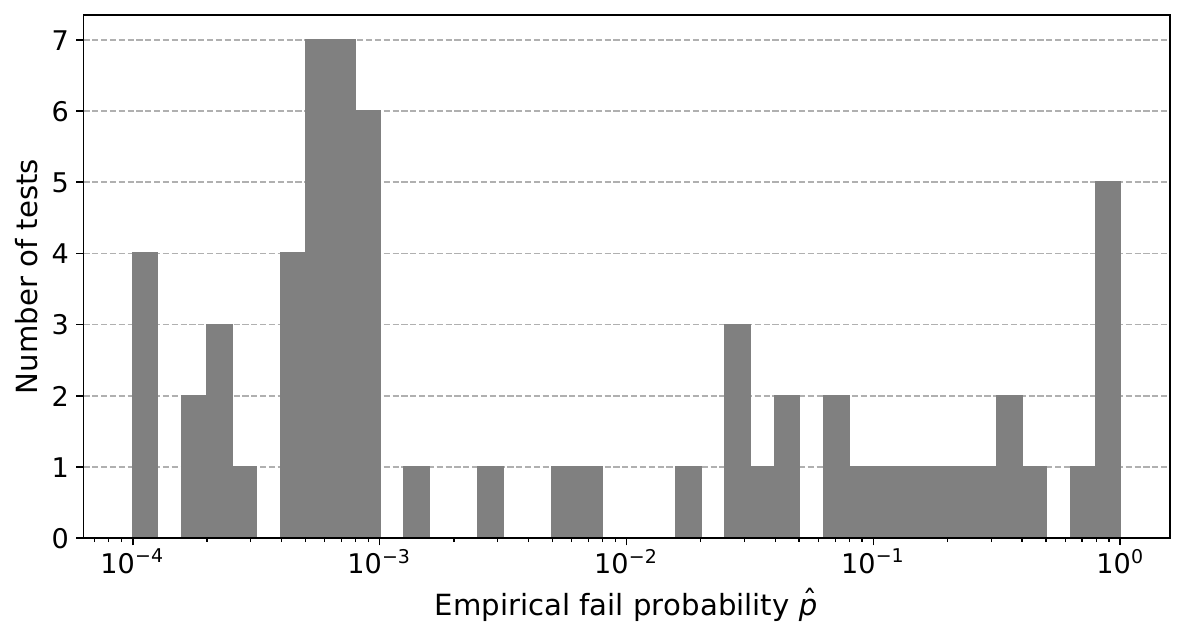}
     \caption{Distribution of empirical failure probabilities $\hat p$ for all flaky tests in our dataset (62 unique tests, aggregated across releases). The distribution is fat-tailed and exhibits two distinct modes: a dominant cluster of low-probability tests concentrated between $\hat p = 10^{-4}$ and $10^{-3}$ (median $\hat p \approx 9\times10^{-4}$; 4~tests at the observable floor of $\hat p = 10^{-4}$), and a smaller cluster of high-probability tests near $\hat p \in [10^{-1}, 1]$, with the most extreme case approaching $\hat p = 1.0$.}

    \label{fig:distribution}
\end{figure}

\Cref{fig:distribution} shows the distribution of these estimates. The distribution is bimodal, comprising a dominant cluster of low-probability tests and a smaller cluster of high-probability tests. The majority of flaky tests concentrate at low failure probabilities: 34 of 62 tests (54.84\%) have $\hat p \le 10^{-3}$, including 4 tests at the observable floor of $\hat p = 10^{-4}$, with a median of $\hat p = 9 \times 10^{-4}$. At the same time, the upper tail is non-negligible: 13 tests exhibit $\hat p \in (0.1, 1)$, and 3 tests fail in nearly every run ($\hat p \ge 0.99$), with the most extreme case approaching $\hat p = 1.0$.

It is easy to miss test flakiness when $\hat{p}$ is low (e.g. $\hat{p} = 10^{-4}$) under small rerun budgets. Let us perform a theoretical analysis to quantify the hardness.

\subsubsection{Theoretical analysis}\label{sec:theoretical_true_p}
How many repetitions are needed to confidently detect a flaky test? Assuming IID test outcomes, failures follow a binomial distribution. To bound the true failure probability $p$ given an empirical estimate $\hat{p}$ at a chosen confidence level, we use the Wilson confidence interval for binomial proportions~\citep{wilson1927probable}; see~\citet{brown2001interval} for a review.

The IID assumption provides a baseline detectability model rather than a guarantee for every execution environment. Repeated HPC executions may be correlated because of shared machine state, filesystem cache warm-up, CPU-frequency scaling, scheduler behavior, dependency resolution, or job contention. Positive temporal or within-batch correlation can cluster outcomes, produce overdispersion relative to a binomial model, and reduce the effective information in 10,000 executions. In that setting, an ordinary Wilson interval may be too narrow and may not attain its nominal coverage. If the conditional failure probability changes across nodes, workloads, or time windows, a single $p$ is an average over the sampled environment rather than an invariant property of the test.

Dependence also invalidates the binomial expression $1-(1-p)^n$ for the probability of observing at least one failure. In particular, clustered failures can create longer failure-free stretches than the independent model predicts, making an IID-derived rerun budget optimistic; other dependence structures can affect the calculation differently. We therefore interpret the Wilson intervals and derived budgets as idealized IID baselines, especially for very small empirical probabilities, rather than guaranteed budgets for every CI or HPC environment. Environment-specific use should first assess temporal and batch effects and, when necessary, use cluster-aware, overdispersed, stratified, or resampling-based uncertainty estimates.

\begin{figure}[tbh]
    \centering
    \includegraphics[width=0.8\linewidth]{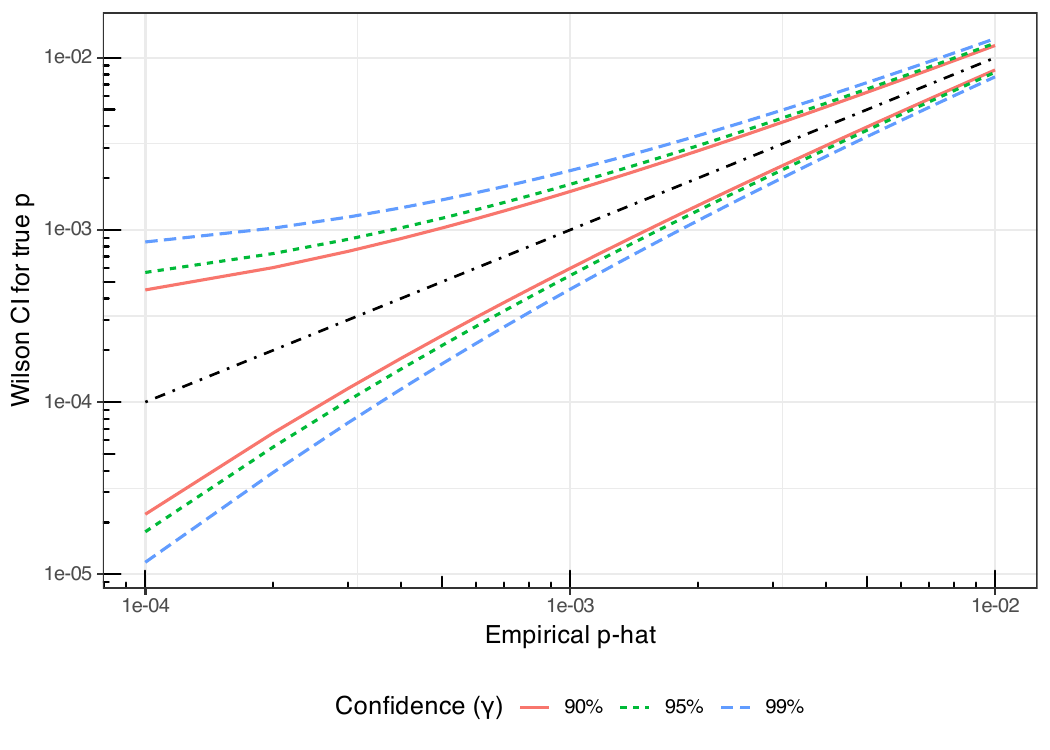}
    \caption{
    Wilson confidence intervals for the true failure probability $p$ as a function of the empirical estimate $\hat{p}$ when $N=10000$. For each confidence level $\gamma \in \{90\%, 95\%, 99\%\}$, the upper and lower bounds are shown as paired lines with matching color and line type. The reference line $p=\hat{p}$ is shown with a dot-dash pattern.}
    \label{fig:wilson_ci}
\end{figure}

\Cref{fig:wilson_ci} shows how the Wilson interval varies with $\hat{p}$ at different confidence levels $\gamma$. As $\hat{p}$ decreases, uncertainty grows relative to the estimated probability.

We next connect these bounds to a rerun budget that is conservative within the IID binomial model. Let $X$ be the number of observed failures after $n$ executions, so that $X \sim \mathrm{Binomial}(n, p)$. In general, the probability of observing exactly $k$ failures is
$$
\Pr(X = k) = \binom{n}{k} p^{k} (1-p)^{n-k}, \qquad k=0,1,\dots,n.
$$
Therefore, the probability of detecting flakiness by observing at least one failure is
\begin{equation}\label{eq:ge_1}
\Pr(X \ge 1) = 1 - \Pr(X = 0)
= 1 - \binom{n}{0} p^0 (1-p)^n
= 1 - (1-p)^n .
\end{equation}
Given a target detection probability $q\in(0,1)$, we require
\begin{equation}\label{eq:n}
1-(1-p)^n \ge q
\quad \Leftrightarrow\quad
(1-p)^n \le 1-q
\quad \Rightarrow\quad
n \ge \frac{\ln(1-q)}{\ln(1-p)},
\end{equation}
as $\ln(1-p)<0$.

In practice, $p$ is unknown. Let $[L,U]$ be the Wilson confidence interval for $p$ with nominal coverage $\gamma$, computed from $N$ runs, and let $L = L_\gamma(\hat{p},N)$ denote its lower bound.
Using $L$ as a conservative lower bound on the unknown $p$, and noting that \Cref{eq:ge_1} is monotonically increasing in $p$, the worst case occurs at $p=L$.
Therefore, a rerun budget that is conservative relative to the Wilson lower bound within the IID model and targets detection probability at least $q$ is
$$
n_{q,\gamma}(\hat p,N)=\left\lceil \frac{\ln(1-q)}{\ln\left[1-L_{\gamma}(\hat p,N)\right]} \right\rceil,
$$
where $\lceil\cdot\rceil$ denotes the ceiling function.

\begin{figure}[tbh]
\centering
\includegraphics[width=0.8\linewidth]{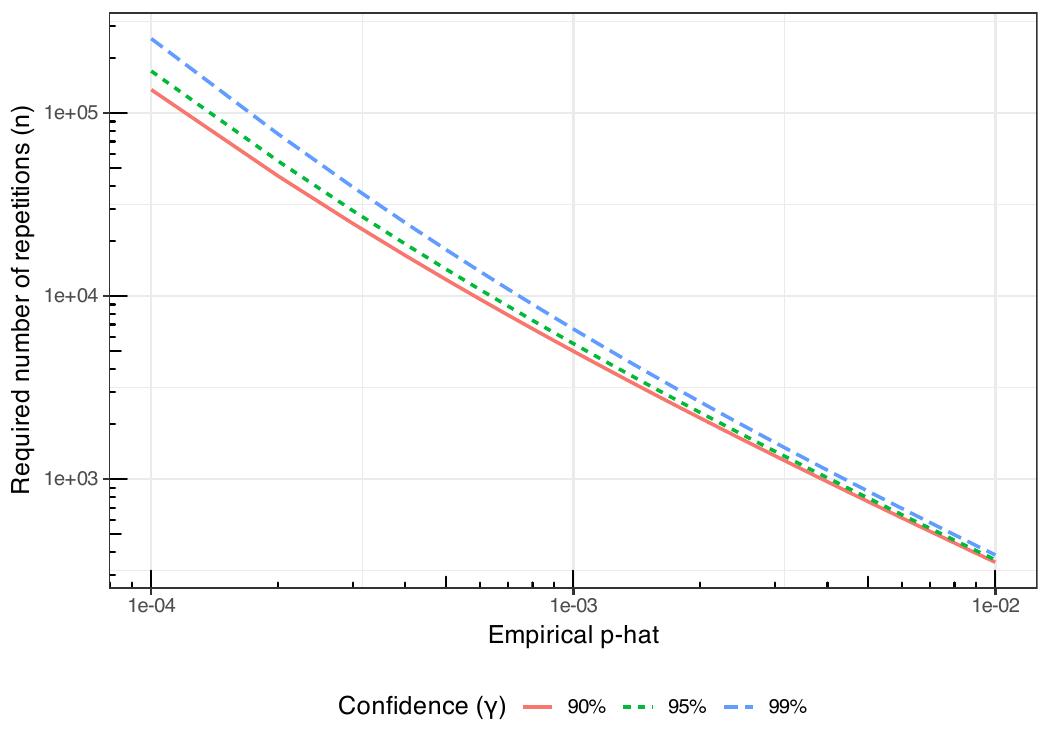}
\caption{
IID-baseline rerun budget for flakiness detection using Wilson lower confidence bounds. The $x$-axis shows the empirical estimate $\hat p$ from $N=10{,}000$ observed runs, and the $y$-axis shows the required number of repetitions $n_{q,\gamma}(\hat p, N)$ with $q=95\%$ and $\gamma\in\{90\%,95\%,99\%\}$. Within the IID model, higher confidence levels require more repetitions.
}
\label{fig:wilson_required_repetitions}
\end{figure}

\Cref{fig:wilson_required_repetitions} plots $n_{q,\gamma}(\hat p,N)$ versus $\hat p$ (for fixed $N$ and $q$), illustrating how required repetitions grow rapidly as $\hat p$ decreases and increase further with higher confidence $\gamma$.
For example, with $q=0.95$ and $\gamma=0.95$, when $N=10000$ and $\hat p=10^{-4}$, then the 95\% Wilson lower bound is $L_{0.95}\approx 1.77\times 10^{-5}$. Plugging these values into \Cref{eq:n}, yields $n_{0.95,0.95}(0.0001, 10000)\approx 1.69\times 10^5$ repetitions. When $\hat p=0.01$ and $L\approx 8.23\times 10^{-3}$, this reduces to $n_{0.95,0.95}(0.01, 10000)\approx 363$ repetitions.

To make these IID-baseline implications concrete for practitioners, consider per-commit CI rerun budgets of $n=50$, $100$, and $1{,}000$. Using $\Pr(X \ge 1) = 1-(1-p)^n$, a test with true failure probability $p=10^{-3}$ is detected with probability approximately $4.9\%$, $9.5\%$, and $63.2\%$, respectively, while a test with $p=10^{-4}$ is detected with probability approximately $0.5\%$, $1.0\%$, and $9.5\%$. In other words, even a process that reruns a test one thousand times has less than a $10\%$ chance, under IID, of surfacing a flake whose true failure probability is on the order of $10^{-4}$. This is consistent with our observation that many flaky \terra tests recur across releases without being caught in every release. Our results do not imply that every test should be executed thousands of times for every pull request. Instead, they motivate a tiered execution strategy: small rerun budgets for routine per-commit validation, intermediate budgets for nightly testing, and larger rerun budgets for pre-release validation or historically flaky tests. Without such tiering, the low-probability mode identified in \Cref{fig:distribution} will remain difficult to detect.

\subsection{RQ4: \RQFour}

\begin{figure}[h!]
  \centering
  \includegraphics[width=\textwidth]{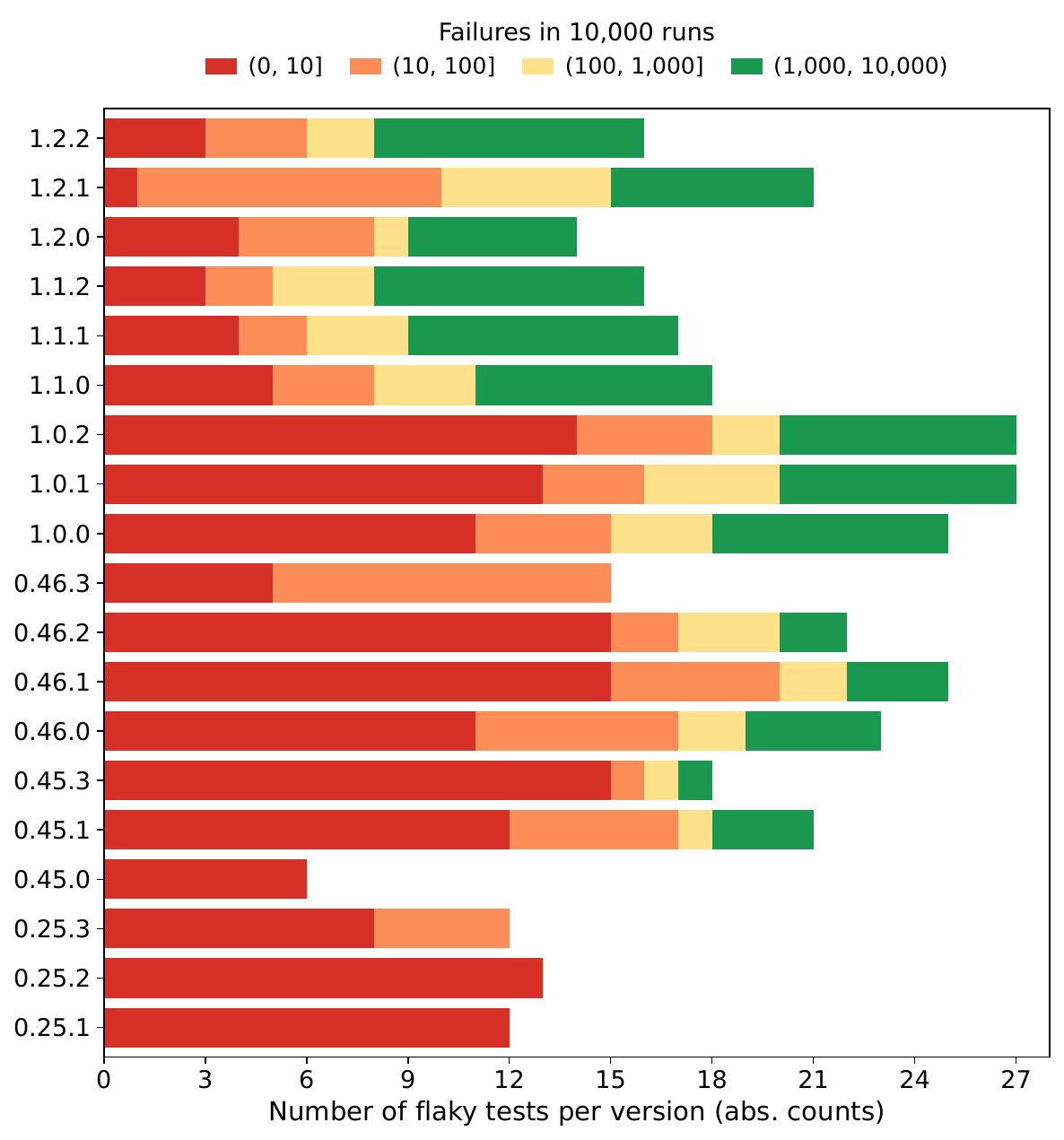}
  \caption{Distribution of failure frequencies for flaky tests across \terra releases. Each horizontal bar partitions a release’s flaky tests into four bins based on the number of failures observed in 10,000 executions.}
  \label{fig:version_flaky_distribution}
\end{figure}

To characterize failure intensity within a release, we grouped empirical failure counts (out of 10,000 runs) into four bins: $(0, 10]$, $(10, 100]$, $(100, 1000]$, and $(1000, 10000]$. As shown in Figure~\ref{fig:version_flaky_distribution}, the distribution is dominated by the lowest bin, where $\hat{p} \leq 10^{-3}$. High-frequency flaky tests ($\hat{p} > 0.1$) are rare but present. Notably, tests in the $(100, 1000]$ bin appear relatively late, starting with v.0.45.1.

\subsection{RQ5: \RQFive}

\begin{figure}[tbp]
    \centering
    \includegraphics[width=0.7\textwidth]{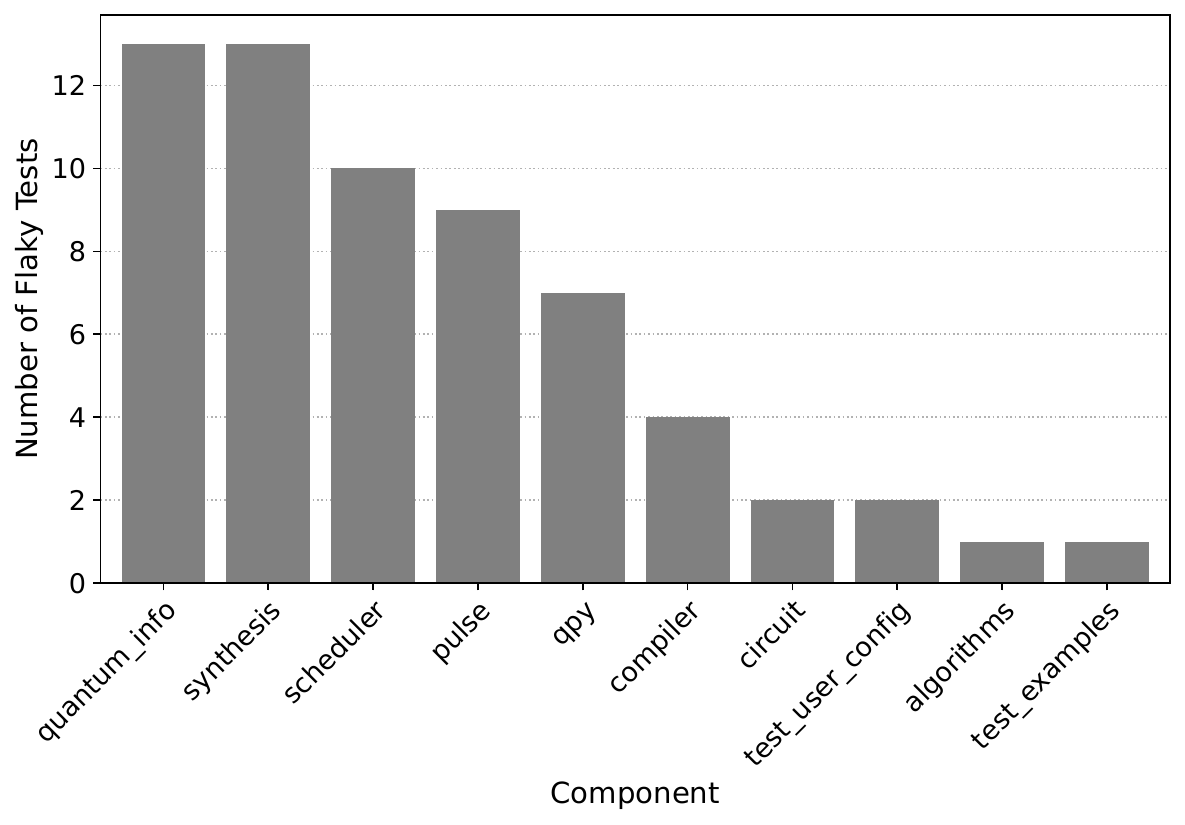}
     \caption{Total number of distinct flaky tests by \terra subcomponent, aggregated across all releases. Subcomponents are sorted in descending order. }
    \label{fig:Number_of_flaky_test_by_component}
\end{figure}

\begin{figure}[tbp]
    \centering
    \includegraphics[width=\textwidth]{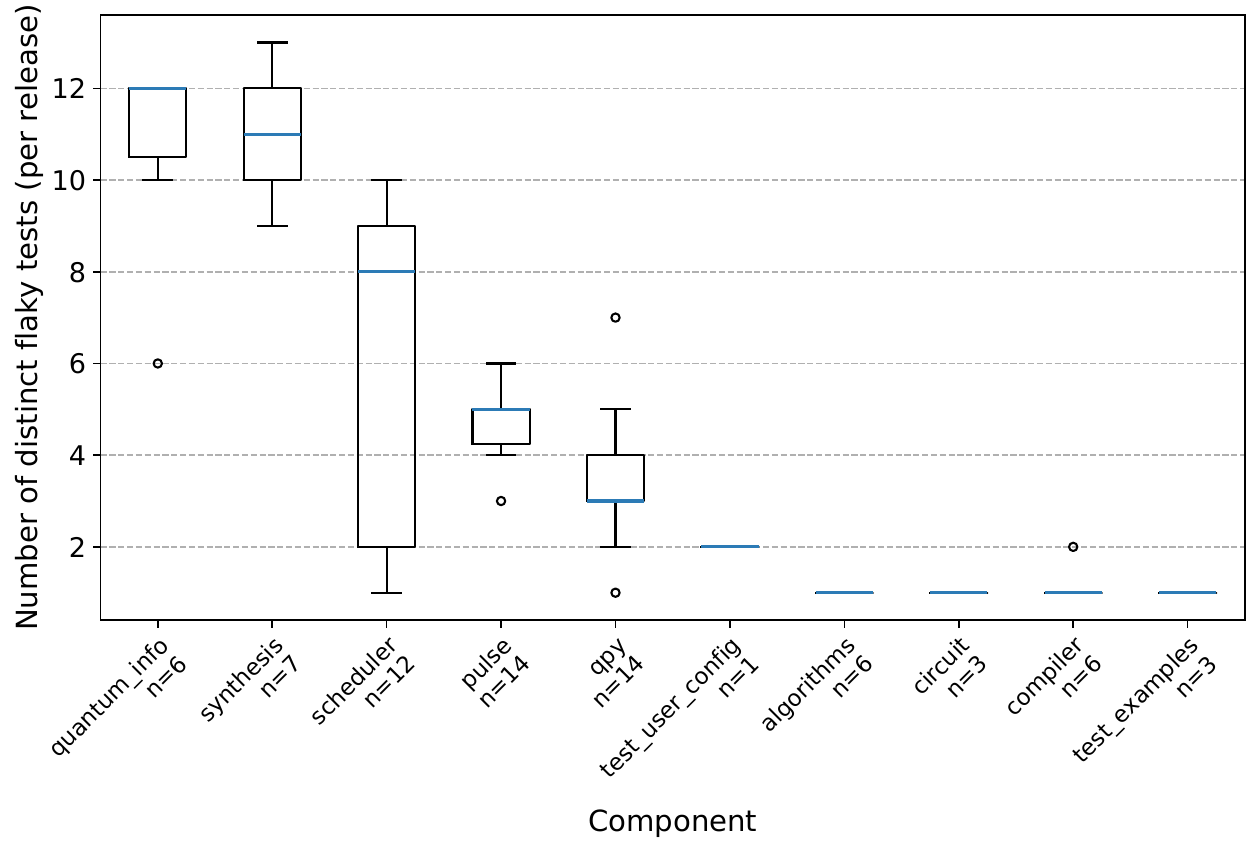}
    \caption{Distribution of flaky tests across \terra subcomponents over all analyzed releases.
    For each subcomponent, the boxplot summarizes the per-release count of distinct flaky test cases, and subcomponents are ordered as in \Cref{fig:Number_of_flaky_test_by_component}, where $n$ indicates the number of releases in which flaky tests are observed for that subcomponent.}
    \label{fig:component_level}
\end{figure}

To understand which parts of \terra are most affected by nondeterminism, we mapped each test to its subcomponent. Subcomponent descriptions are provided in \Cref{app:modules}. \Cref{fig:component_level,fig:Number_of_flaky_test_by_component} show that flakiness is highly nonuniform across the subcomponents.

\Cref{fig:Number_of_flaky_test_by_component} indicates that the \texttt{quantum\_info} and \texttt{synthesis} subcomponents have the highest number of distinct flaky tests (13), followed by \texttt{scheduler} (10) and \texttt{pulse} (9). In contrast, subcomponents such as \texttt{algorithms} and \texttt{text\_examples} each contain only one flaky test.

\Cref{fig:component_level} provides a per-release breakdown, highlighting further variation across subcomponents. For example, the \texttt{quantum\_info} subcomponent shows the highest median number of distinct flaky tests per release (median $\approx 12$, $n=6$ releases), followed closely by \texttt{synthesis} (median $\approx 11$, $n=7$).
The \texttt{scheduler} component, while present in 10 releases, exhibits the widest spread---ranging from 1 to 10 flaky tests per release---suggesting high sensitivity to release-to-release changes. The \texttt{pulse} and \texttt{qpy} subcomponents appear in the most releases ($n=14$ each), yet maintain a narrower range of flaky test counts, with \texttt{pulse} having a median of 5 and \texttt{qpy} a median of 3.

Overall, the flakiness observed under our protocol is concentrated in a few \terra subcomponents, such as \texttt{quantum\_info} and \texttt{synthesis}. The wide variation in flaky-test frequency across releases also indicates that some components may be more sensitive to changes in the codebase or external dependencies. Understanding this distribution may help prioritize stabilization efforts and guide targeted improvements in test reliability.

The component totals should not be interpreted as counts of independent underlying defects. Our exploratory manual analysis indicates that some high counts arise from families of tests that exercise the same low-level implementation. For example, several precision-sensitive tests using \texttt{TwoQubitWeylDecomposition} were located in \texttt{quantum\_info} and later moved to \texttt{synthesis}; small numerical differences near machine precision could cross strict assertion tolerances. This shared mechanism helps explain why both components contain many observed flaky tests. Some transpiler failures instead involve stochastic compilation decisions: in one manually verified case, unseeded layout and routing on a disconnected target produced inconsistent outcomes and the fix was to set \texttt{seed\_transpiler}. These examples suggest that component concentrations reflect a mixture of shared numerical kernels, related test families, code reorganization, and exposure to stochastic algorithms. Because this mapping is observational and manually reconstructed, we treat these explanations as qualitative hypotheses rather than complete causal accounts.

\subsection{RQ6: \RQSix}\label{sec:individual_test_case_behaviour}

We analyzed time-series patterns of flaky appearances across releases and identified three characteristic temporal profiles.

\begin{figure}[tbp]
    \centering
    \begin{subfigure}{0.99\textwidth}
        \includegraphics[width=\textwidth, page=1]{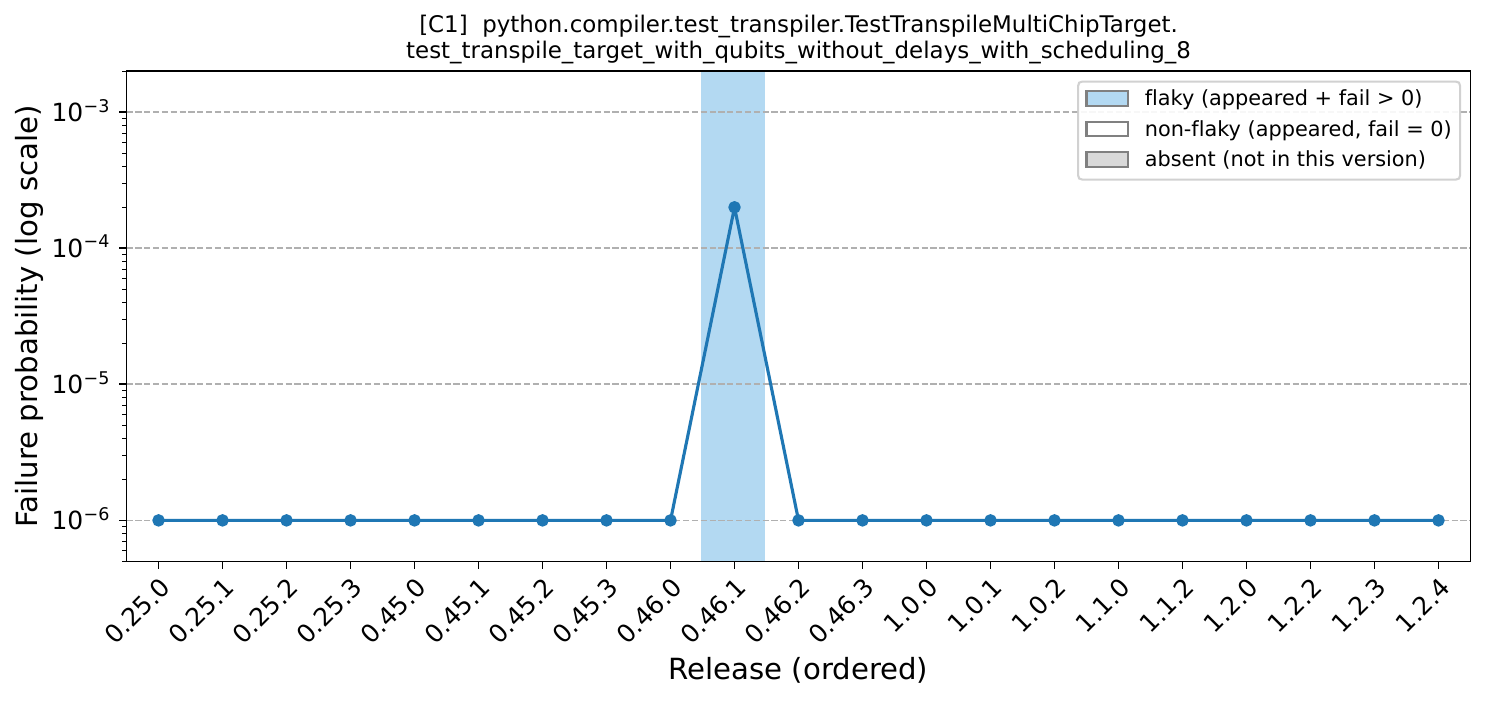}
    \end{subfigure}
    \par\medskip
    \begin{subfigure}{0.99\textwidth}
        \includegraphics[width=\textwidth, page=2]{C1_bigfont.pdf}
    \end{subfigure}
   \caption{\COne pattern. The temporal pattern in the right panel does not imply a sequence of flaky–fixed–reintroduced states. Rather, the underlying failure event is extremely rare, so the true failure probability is low and may not be consistently observed (see \Cref{sec:theoretical_true_p}). Because a log scale cannot display zero, we use $10^{-6}$ as a placeholder for zero observed failures. The lowest observable non-zero rate is $10^{-4}$, corresponding to one failure in 10,000 runs.}
    \label{fig:C1}
\end{figure}

\begin{figure}[tbp]
    \centering
    \begin{subfigure}{0.99\textwidth}
        \includegraphics[width=\textwidth, page=1]{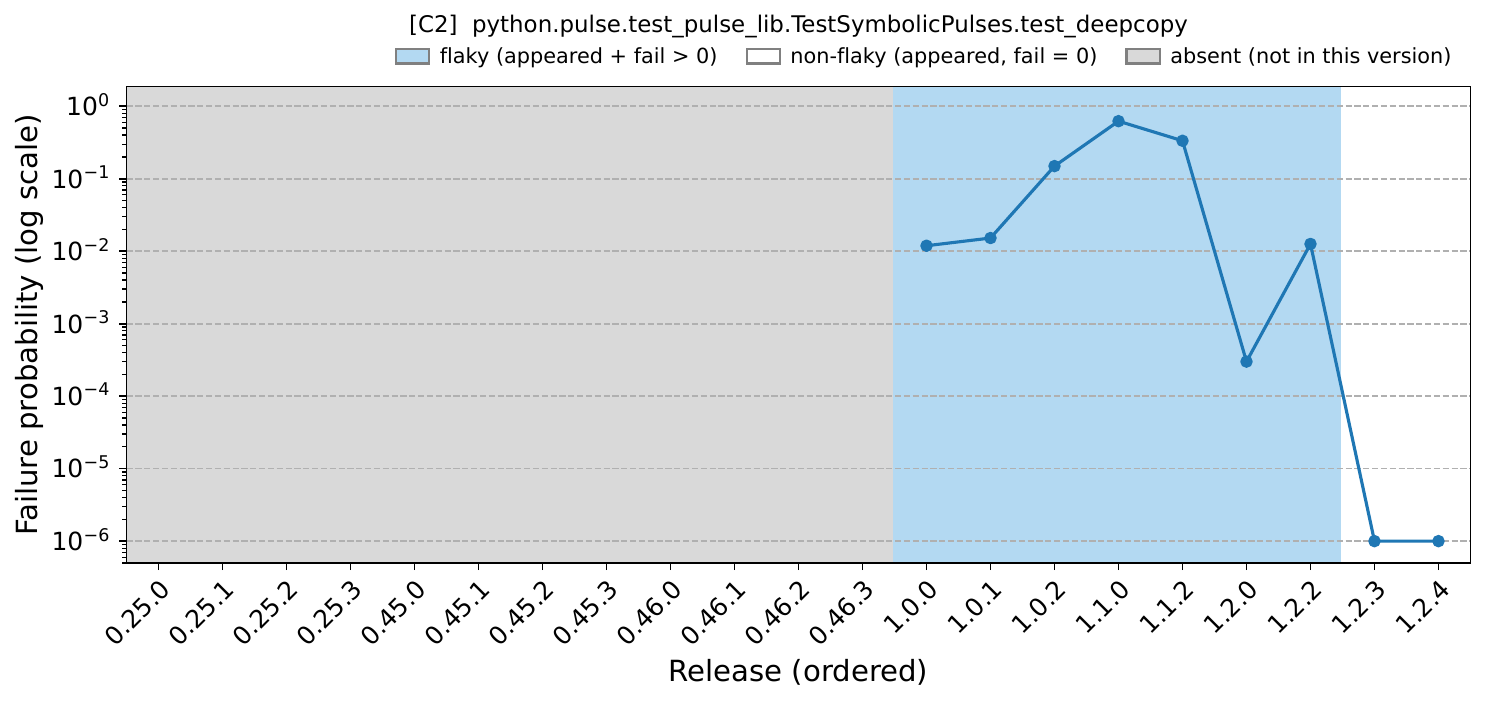}
    \end{subfigure}
    \par\medskip
    \begin{subfigure}{0.99\textwidth}
        \includegraphics[width=\textwidth, page=2]{C2_bigfont.pdf}
    \end{subfigure}
    \caption{\CTwo pattern. The gray background marks releases in which
    the test case did not exist or was skipped. In the left panel, the
    test is present across several releases and consistently flaky until
    removal. In the right panel, the test is introduced later, remains
    consistently flaky, and appears to be fixed only in the final two
    releases. As log scales cannot display zero, we use $10^{-6}$ for
    zero observed failures; the lowest observable non-zero rate is $10^{-4}$.}
    \label{fig:C2}
\end{figure}

\begin{figure}[tbp]
    \centering
    \begin{subfigure}{0.99\textwidth}
        \includegraphics[width=\textwidth, page=1]{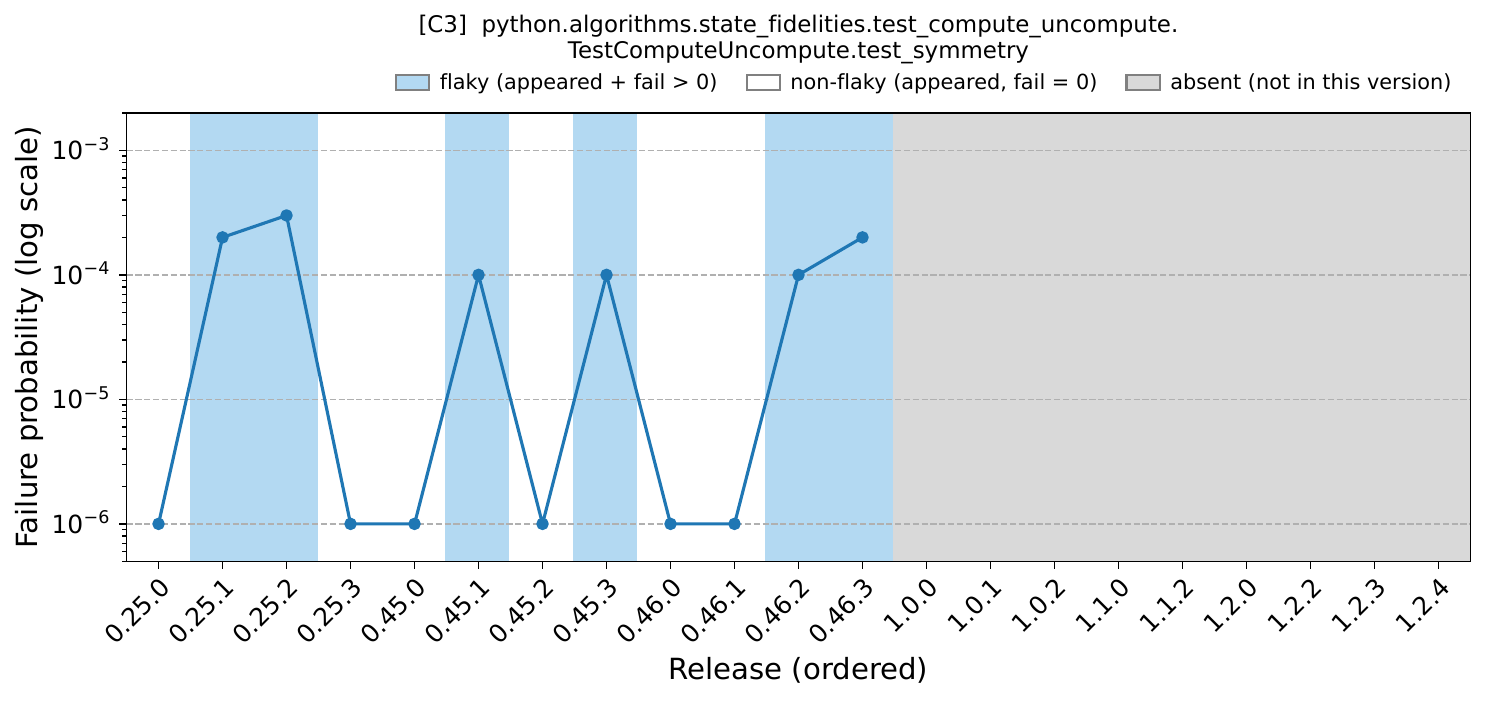}
    \end{subfigure}
    \par\medskip
    \begin{subfigure}{0.99\textwidth}
        \includegraphics[width=\textwidth, page=2]{C3_bigfont.pdf}
    \end{subfigure}
   \caption{\CThree pattern. These examples show test cases that intermittently alternate between flaky and stable states. We do not speculate on the underlying causes, but this pattern contrasts with consistently failing tests such as those in \Cref{fig:C2}. As in \Cref{fig:C1,fig:C2}, $10^{-6}$ denotes zero observed failures, and the smallest observable non-zero rate is $10^{-4}$.}
    \label{fig:C3}
\end{figure}

\begin{description}[font=\normalfont\bfseries]
    \item[\COne] Tests flaky in $\le 15\%$ of their active releases. This subtype accounts for 14 tests (22.58\%). These failures are rare; examples are presented in \Cref{fig:C1}.

    \item[\CTwo] Tests flaky in $\ge 70\%$ of the releases in which they appear. Only 7 tests (11.29\%) fall into this group. They fail with variable intensity and are often associated with unstable transformations in the transpiler; examples are presented in \Cref{fig:C2}.

    \item[\CThree] Tests that alternate between flaky and stable states across multiple releases.
    This subtype includes 41 tests (66.13\%) and is characterized by an intermittent, back-and-forth pattern where each flaky period is brief. Examples are presented in \Cref{fig:C3}.

\end{description}

\begin{figure}[ht!]
    \centering
    \includegraphics[width=0.99\textwidth]
    {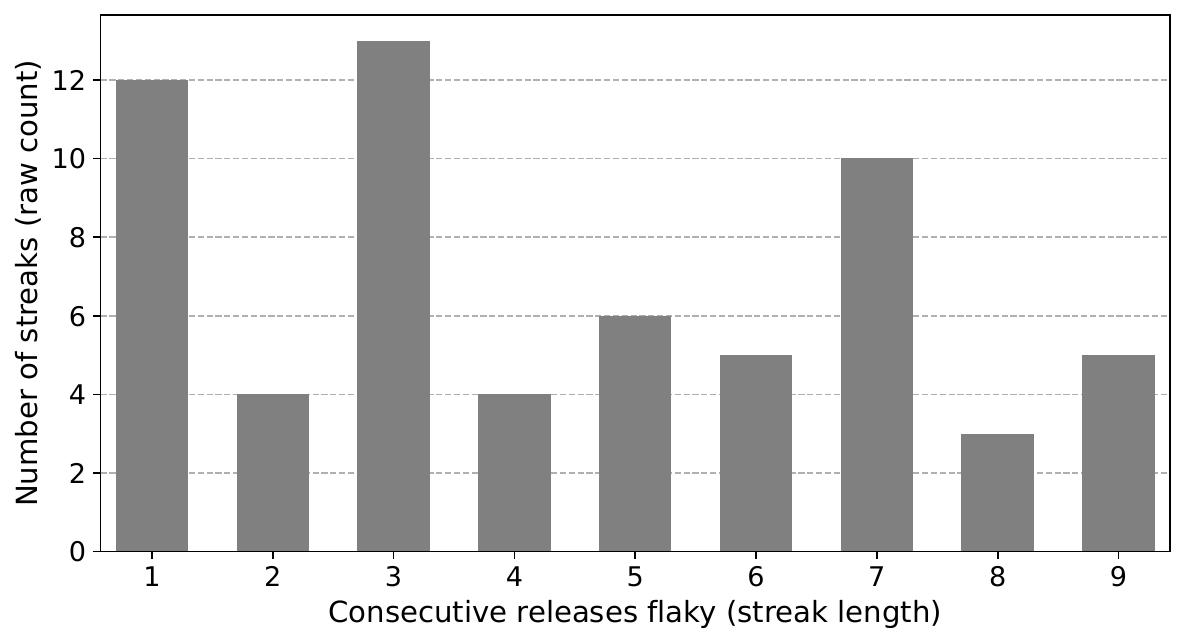}
    \caption{Longest consecutive flaky-release streak per test across \terra releases. Bars show the number of flaky tests whose maximum streak length is $m$, where a streak is a maximal run of $m$ consecutive releases in which a test is flaky. Streak lengths span the full observed range ($1$--$9$), and the most frequent maximum streak length is $m=3$ (13 tests), with $m=1$ (12 tests) and $m=7$ (10 tests) also common.}
    \label{fig:flaky_streaks}
\end{figure}

The streak-length distribution in Figure~\ref{fig:flaky_streaks} shows that maximum consecutive-flakiness streaks span the full observed range of $1$--$9$ releases and are broadly distributed across it. The most frequent maximum streak is $m=3$ (13 tests), closely followed by isolated single-release flakiness ($m=1$, 12 tests) and by long streaks of $m=7$ (10 tests).
Only 12 of 62 tests (19.35\%) have a maximum consecutive flaky-release streak of one release, although some of these tests may recur in multiple non-consecutive releases. We also observe 18 tests (29.03\%) sustain streaks of 7 or more consecutive releases. Flakiness in \terra is therefore neither purely fragmented nor uniformly persistent: once a test becomes flaky, it frequently---though not always---remains flaky across several consecutive releases.

\subsection{Exploratory Root-Cause Categories and Fix Patterns} \label{sec:root-cause-results}
To explore why the flaky tests in our dataset fail and how they were subsequently addressed in Qiskit, we manually inspected each of the 62 unique flaky tests and assigned a suspected root-cause category and an observed fix-pattern category. We use the taxonomy for quantum flaky tests proposed by~\citet{sivaloganathan2026automating}, which builds on the classical flaky-test taxonomies of~\citet{luo2014empirical,lam2019root}. For clarity, we group low-count or ambiguous categories into an \emph{Others} bucket. These labels are evidence-consistent interpretations rather than definitive causal diagnoses.
The full per-test annotations, including the specific root-cause and fix subcategories, are provided in the supplementary spreadsheet \texttt{qiskit\_flaky\_tests\_root\_cause\_fix\_labels.xlsx}, released as part of the public dataset at \url{https://zenodo.org/records/20500432}.

\begin{table*}[t]
\centering
\caption{Exploratory counts of suspected root-cause categories and observed fix patterns based on manual verification.}
\label{tab:manual_root_cause_fix_summary}
\resizebox{\textwidth}{!}{
\begin{tabular}{@{}l|rrrrr|rr@{}}
\toprule
                     & \multicolumn{5}{c|}{Fix category}
                     &              &              \\
\cmidrule(lr){2-6}
Root-cause category   & Remove Functionality
                     & Rewrite Component
                     & Fix Seed
                     & Change Test Env.
                     & Remove Test Case
                     & Grand Total
                     & Percentage \\
\midrule
Floating Point Operations & 6  & 24 & 0 & 0 & 0 & 30 & 48.4\% \\
Multi-threading           & 0  & 0  & 0 & 0 & 1 & 1  & 1.6\%  \\
Randomness                & 0  & 0  & 3 & 0 & 1 & 4  & 6.5\%  \\
Others                    & 23 & 1  & 0 & 3 & 0 & 27 & 43.5\% \\
\midrule
Grand Total               & 29 & 25 & 3 & 3 & 2 & 62 & 100\% \\
Percentage                & 46.8\% & 40.3\% & 4.8\% & 4.8\% & 3.2\% & 100\% & \\
\bottomrule
\end{tabular}
}
\end{table*}

\subsubsection{Verification Method}
For each unique flaky test, two authors independently labeled the suspected root cause and fix pattern. Each reviewer examined the same evidence sources without seeing the other's labels: (i) the test source at the last release in which the test was flaky, (ii) the failing assertion and any reproducible traceback, (iii) the implementation of the exercised component, such as a transpiler pass, a synthesis routine, or the symbolic-pulse backend, and (iv) the version-control history between the last flaky release and the first stable or removal release, including commits, pull requests, and linked GitHub issues when available.

The reviewers then reconciled their labels jointly. Agreed labels were retained, and disagreements were resolved by re-examining the evidence until consensus was reached. We also used the large language model Claude Opus 4.7~\citep{anthropic2025claude} to cross-check the reviewers' notes against the same evidence and flag possible inconsistencies; however, the final labels were always assigned by the reviewers. The reconciled labels form the basis for Table~\ref{tab:manual_root_cause_fix_summary}.

This process reduces the risk of individual error but does not establish that the category set is complete or that another team would reproduce every reconciled label. Repository history may show that a failure disappeared after a change without proving why that change had the observed effect. We therefore treat the categories as exploratory, best-effort classifications and avoid using them to make population-level causal claims. The large \emph{Others} group further indicates that the summary taxonomy does not provide a specific explanation for every observed test; the supplementary spreadsheet preserves the more detailed per-test annotations.

\subsubsection{Suspected Root-Cause Category}
We classify suspected mechanisms using the quantum flaky-test taxonomy of~\citet{sivaloganathan2026automating}, which defines nine root-cause categories. Four summary categories occur in our data: \emph{Floating Point Operations}, \emph{Randomness}, \emph{Multi-Threading}, and \emph{Others}. The remaining five categories (\emph{Software Environment}, \emph{Visualization}, \emph{Unhandled Exception}, \emph{Network}, and \emph{Unordered Collection}) did not match any of our observed flaky tests and are therefore omitted from the discussion below.

\paragraph{Floating Point Operations} This category covers tests for which the evidence points to numerical precision issues, such as round-off errors. It is the dominant suspected mechanism in our dataset, accounting for 30 of 62 tests ($48.4\%$). The recurring pattern is a strict equality or tight-tolerance assertion applied to values computed through quantum operators or transformations, where small differences near machine precision can change the test outcome. A representative example is the family of tests in \texttt{quantum\_info} (later moved to \texttt{synthesis}) using \texttt{TwoQubitWeylDecomposition}. In these tests, a round-trip through the Weyl/KAK decomposition produced matrix entries that differed by approximately $1.75\times10^{-16}$ between runs and violated the assertion. The underlying issue (Qiskit \#11676) was ultimately addressed by reimplementing the decomposition kernel in Rust (PR~\#11946). The much higher share of floating-point classifications in our dynamic dataset than the $9.6\%$ reported from issue and pull-request text in~\citet{sivaloganathan2026automating} is consistent with the possibility that dynamic re-execution surfaces precision-driven failures that may never generate an explicit issue report.

\paragraph{Randomness} This category covers tests for which pseudo-random choices vary across runs when no fixed seed is used. It accounts for 4 of 62 tests ($6.5\%$). A representative case is \texttt{test\_transpile\_target\_with\_qubits\_without\_delays\_with\_scheduling\_8} in the transpiler subcomponent, where unseeded layout and routing search on a disconnected multi-chip target produced nondeterministic outcomes. The fix was to set \texttt{seed\_transpiler}. Randomness is the dominant category ($19.2\%$) in the issue and pull-request data of~\citet{sivaloganathan2026automating}, but it represents only a small share of our dynamically observed failures. One plausible explanation is that our analysis uses code snapshots from shipped Qiskit releases, so seed-controllable randomness may have been identified and fixed during release preparation, leaving fewer such failures to be observed through re-execution.

\paragraph{Multi-Threading} This category covers tests for which the evidence points to concurrency, scheduling, or overload effects. In our dataset, only 1 of 62 tests ($1.6\%$) falls into this category, substantially lower than the $13.7\%$ reported in the issue and pull-request data of~\citet{sivaloganathan2026automating}.

\paragraph{Others}
The remaining 27 tests ($43.5\%$) fall into the \emph{Others} bucket. These tests correspond to low-count or ambiguous mechanisms that did not warrant separate categories in the summary table. The size of this group reinforces the exploratory and incomplete nature of the summary classification. Detailed per-test labels are provided in the supplementary spreadsheet.

\subsubsection{Fix Category}
We also classify how each flaky test was ultimately addressed in the Qiskit codebase. The observed fix patterns only partially overlap with the taxonomy of~\citet{sivaloganathan2026automating}: \emph{Fix Seed} maps directly to controlling randomness, and \emph{Change Test Environment} is closely related to altering the software or test environment. The remaining patterns (\emph{Remove Functionality}, \emph{Rewrite Component}, and \emph{Remove Test Case}) emerged from our manual inspection of Qiskit's version history. We define these categories below.

\paragraph{Remove Functionality}
This category captures cases where the flaky test disappeared because the functionality it exercised was removed, deprecated, or subsumed by another implementation. It is the most common fix pattern in our data, accounting for 29 of 62 tests ($46.8\%$). In these cases, the disappearance of flakiness does not necessarily mean that the underlying nondeterminism was diagnosed and repaired directly; rather, the affected functionality was no longer part of the maintained code path. Thus, some flaky tests appear to have been resolved as a by-product of broader codebase evolution rather than through targeted flaky-test repair.

\paragraph{Rewrite Component}
This category captures cases where the exercised implementation was substantially rewritten while preserving the intended functionality. It is the second most common fix pattern, accounting for 25 of 62 tests ($40.3\%$). Several floating-point-related failures disappeared after numerical or synthesis components were reimplemented, including cases where Python implementations were replaced by Rust-based kernels. For example, the \texttt{TwoQubitWeylDecomposition} failures discussed above were addressed by reimplementing the decomposition kernel in Rust. We do not claim that the language change alone explains the disappearance of flakiness; rather, the rewrite changed the numerical implementation sufficiently that the previously observed precision-sensitive failures no longer manifested.

\paragraph{Fix Seed}
This category covers cases where nondeterminism was addressed by controlling a random source, typically by setting or propagating a deterministic seed. It accounts for 3 of 62 tests ($4.8\%$) and corresponds closely to the \emph{Randomness} root-cause category. For example, tests involving stochastic transpiler behavior can fail when layout, routing, or synthesis choices vary across executions while the assertion implicitly assumes one outcome. Setting the seed makes the test deterministic by forcing the same randomized choice across runs.

\paragraph{Change Test Environment}
This category captures cases where flakiness was addressed by changing the test environment rather than the tested functionality. It accounts for 3 of 62 tests ($4.8\%$). Such changes may involve test configuration, dependencies, execution conditions, or other environmental assumptions that affect determinism. These cases are close to the \emph{Alter Software Environment} fix pattern in prior taxonomies; we use the narrower label \emph{Change Test Environment} to emphasize that the observed change affected the testing setup rather than necessarily changing the production implementation.

\paragraph{Remove Test Case}
This category captures cases where the flaky test itself was removed while the broader functionality was not clearly removed at the same time. It accounts for 2 of 62 tests ($3.2\%$). Unlike \emph{Remove Functionality}, this pattern indicates that the unstable test was eliminated from the suite even though the exercised functionality may have remained. As with functionality removal, this does not necessarily mean that the root cause of nondeterminism was repaired.

Overall, the fix-pattern distribution shows that most flaky tests in our dataset were not associated with small, localized changes such as setting a seed or adjusting the test environment. Instead, 54 of 62 tests ($87.1\%$) disappeared after either functionality removal or substantial component rewrites. This finding should be interpreted cautiously: version-control history reveals what changed, but not always why. A test not being fixed directly should not be read as evidence that developers ignored or deprioritized it. In some cases, the affected component may have already been slated for removal, deprecation, or rewrite, making a targeted flaky-test repair economically unjustified. Leaving the test unresolved until the broader component change landed may therefore have been a deliberate maintenance trade-off rather than a lack of attention to test quality. Because we cannot generally observe developers' internal rationale from repository history alone, we treat this explanation as an interpretation rather than a verified cause. The observed association nevertheless suggests that flaky-test disappearance in this dataset often coincides with broader software maintenance and architectural evolution, not only with targeted flaky-test debugging.

\section{Discussion}
\label{sec:discussion}

The preceding section presented the empirical findings for each research question. We now synthesize these findings and discuss their implications for testing practice, maintenance, and future tool development, while keeping the case-study scope explicit.

\subsection{Scope and Quantum Specificity}
\label{sec:discussion-scope}

The behaviors studied here (inconsistent outcomes, rare failures, cross-release recurrence, numerical sensitivity, and unseeded randomness) are not inherently unique to quantum software. The tests execute on classical computing infrastructure, and our analysis does not estimate a fraction of failures attributable to intrinsically quantum hardware phenomena. The quantum-software significance lies in the domain context: the failures occur in implementations of circuit synthesis, quantum-information transformations, and topology-constrained compilation, and the study supplies large-scale execution evidence that has previously been scarce in this ecosystem.

This design cannot answer what proportion of \terra's flaky tests exists ``because'' the software is quantum. Such an attribution would require a comparative or intervention-based design that separates domain-specific algorithms from general software mechanisms and, for hardware-facing systems, separates device-level stochasticity from classical orchestration. Our exploratory root-cause analysis instead identifies immediate mechanisms in the observed implementation. These mechanisms can be informative without implying that they would be absent from classical scientific or compiler software.

\subsection{Cross-Cutting Observations}
\label{sec:discussion-crosscut}

Considering the six RQs jointly reveals a ``rare-but-recurrent flakiness'' pattern. Combining the observations in \Cref{tab:count_fraction,tab:test-per-version} and \Cref{fig:distribution}, we observe the coexistence of three properties that, taken together, define a particularly difficult-to-handle class of flakiness in \terra:
\begin{enumerate*}[label=(\roman*)]
    \item per-release flakiness rates are low, ranging from 0\% to 0.17\%;
    \item most flaky tests recur across multiple releases, with 52 of 62 unique flaky tests (83.87\%) appearing in more than one release; and
    \item the empirical failure probabilities of many flaky tests are very small, with a median $\hat{p} \approx 9 \times 10^{-4}$ and 34 tests (54.8\%) at or below $10^{-3}$.
\end{enumerate*}
A test that fails roughly once in every thousand or ten thousand executions is unlikely to be observed under typical CI rerun budgets, yet such a test may continue to appear across releases. We refer to this observation as ``rare-but-recurrent flakiness'': flaky behavior that is infrequent enough to evade ordinary CI rerun budgets, yet recurrent enough to remain relevant across releases.

Rarity within a release and persistence across releases are distinct properties. A failure may have low probability because it requires an uncommon pseudo-random choice, a numerical result near an assertion boundary, a particular interleaving, or a rare environmental state. The same mechanism can recur over many releases if the relevant implementation and test oracle remain in place. In our data, the precision-sensitive \texttt{TwoQubitWeylDecomposition} tests and an unseeded transpiler search provide examples of numerical and stochastic pathways, respectively. These cases offer partial explanations, not a complete causal account of all rare-but-recurrent tests. Identifying the triggering state for each test would require controlled perturbation or intervention beyond the observational rerun design.

\subsection{Implications for Quantum Software Testing Practice}
\label{sec:discussion-practice}

The patterns above translate into several practical recommendations for teams developing or maintaining quantum software stacks of \terra's scale and complexity.

\paragraph{Maintain cross-release flaky-test histories}
Because most flaky tests in our dataset recur across multiple releases, release-local quarantine policies may lose information needed for diagnosis. A test that appears stable in one release may still have a history of intermittent flakiness. We therefore recommend maintaining a persistent registry of flaky test identifiers, together with their empirical failure rates, affected releases, suspected cause category, and applied fix pattern where available. Such a registry would make recurrence patterns explicit during triage and would allow analyses like those in \Cref{sec:individual_test_case_behaviour} to be performed continuously rather than retrospectively.

\paragraph{Triage tests by temporal pattern and failure intensity}
The three temporal modes identified in RQ6 suggest different response strategies. Persistently Flaky tests should be prioritized because they represent recurring reliability signals rather than isolated anomalies. Intermittently Flaky tests should be treated as recurrence-prone and may benefit from environment lock-in, dependency pinning, seed control, and closer inspection of tolerance thresholds or nondeterministic test logic. Rarely Flaky tests should be interpreted cautiously: dynamic outcomes alone may be insufficient to identify the mechanism, and tests that intentionally exercise stochastic behavior may require statistical assertions rather than deterministic pass/fail thresholds.

\paragraph{Treat numerical tolerance and seed control as first-class testing concerns}
Our exploratory root-cause analysis (\Cref{tab:manual_root_cause_fix_summary}) identifies floating-point operations as the dominant suspected mechanism, accounting for 30 of 62 flaky tests (48.4\%), typically involving strict-equality or tight-tolerance assertions on values computed through quantum operators and transformations. This suggests that tolerance thresholds should be chosen deliberately and documented as explicit test-design decisions rather than left implicit. Although randomness accounts for only 4 tests (6.5\%) in our dataset, 3 such cases were associated with seed-control fixes, indicating that propagating deterministic seeds through layout, routing, and synthesis remains a cheap and effective safeguard for the stochastic components of \terra.

\paragraph{Avoid interpreting zero observed failures as proof of resolution}
\Cref{tab:count_fraction} shows zero observed flaky tests in v.0.25.0, v.0.45.2, v.1.2.3, and v.1.2.4. While the absence of observed flakiness in the two most recent releases may reflect genuine stabilization, our probabilistic analysis cautions against interpreting zero observations as proof that flakiness has been eliminated. A test with a very low failure probability can produce zero failures in 10,000 executions with non-negligible probability. Thus, zero observed failures should be read as ``no flakiness detected at this budget,'' not necessarily as ``flakiness eliminated.''

\subsection{Implications for Tool and Research Development}
\label{sec:discussion-research}

Beyond immediate testing practice, our findings inform several directions for tooling and research targeted at quantum software reliability.

\paragraph{Execution-level benchmark data complement existing labels}
Existing machine-learning approaches to quantum flaky-test detection rely largely on incident-reported or pre-existing labeled cases. Our findings suggest that such labels may under-represent low-probability flaky tests, which are unlikely to be noticed under ordinary CI budgets. The dynamic execution data produced in this study provide a complementary source of supervision and evaluation under our protocol: future predictors can be assessed not only by binary flaky/non-flaky labels, but also by empirical failure probability, recurrence across releases, subcomponent membership, and historical fix patterns.

\paragraph{Release-level dynamic evidence captures persistent flaky tests}
Our dynamic analysis characterizes flaky tests that are present at release points, and therefore reflects the subset of flaky behavior that survived the development and integration process. This perspective is complementary to issue- and pull-request-based taxonomies, but it should not be interpreted as a complete picture of all flaky tests encountered during development. Some flaky tests may be easier to detect, reproduce, or fix during regular development and CI, and may therefore be removed before a release is shipped. Conversely, the flaky tests observed in our study may represent more persistent or harder-to-eliminate cases.

This distinction matters when comparing root-cause distributions across studies. Floating-point precision failures account for 48.4\% of the release-level flaky tests in our dataset, while randomness and multi-threading are less frequent. However, this does not imply that precision issues dominate the full population of quantum flaky tests across the development lifecycle. Rather, our results show that release-level dynamic studies and report-level studies can expose different slices of the flaky-test population. Characterizing flaky tests that arise and are resolved between releases remains an important direction for future work.

\paragraph{Replication beyond one quantum framework remains necessary}
Finally, the patterns reported here are derived from one quantum-software framework, Qiskit \terra, under a specific Linux-based containerized test environment. Although \terra is large, mature, and widely used, replication across other quantum stacks such as Cirq, PennyLane, and Q\#, as well as across hardware-backed execution environments, is needed before these patterns can be generalized to quantum software as a whole. Frameworks with compiler and simulator layers similar in broad function to \terra may share concerns involving numerical tolerance, seed propagation, and constrained routing, while different hardware topologies and programming models may produce different algorithms and failure modes. Hardware-backed runtimes and cloud clients may additionally expose network, service, calibration, queueing, and device-related variability that this study does not exercise. The protocol and dataset are intended to support such replication; the empirical rates and category proportions are not assumed to transfer unchanged.

\subsection{Summary of Practical Recommendations}
\label{sec:discussion-recommendations}

Based on the results and discussion above, we recommend that quantum software projects: (1) maintain cross-release histories of flaky tests; (2) use tiered rerun budgets for routine, nightly, pre-release, and historically flaky tests; (3) prioritize triage using both empirical failure probability and temporal recurrence; (4) treat seed control, numerical tolerance, and stochastic assertions as explicit test-design decisions; (5) record suspected root causes and applied fix patterns alongside flaky-test reports; and (6) validate future static or ML-based flaky-test detectors against dynamic execution-level evidence.

Together, these recommendations emphasize that flaky-test management in quantum software should be both probabilistic and historical. Low observed flakiness rates do not necessarily imply low engineering risk when failures are rare, recurrent, and difficult to detect under ordinary CI budgets.

\section{Threats to Validity}\label{sec:threats}

We discuss threats to the validity of our case study following the classifications of \citet{wohlin2012experimentation,yin2009case}.

\subsection{Internal}
Our methodology is derived from the classical experimental methodology of FlakeFlagger~\citep{alshammari2021flakeflagger}: a high-budget rerun protocol (10,000 executions per test per release) in a controlled environment without injecting additional nondeterminism. While this approach provides rich execution-level data, it also introduces several risks.

First, repeated executions on an HPC cluster may not be independent or identically distributed because of factors such as CPU-frequency scaling, filesystem-cache warm-up, scheduler behavior, and job contention. Positive temporal or within-batch correlation can produce overdispersion, reduce the effective information in the sample, and make an ordinary Wilson interval too narrow. Nonstationarity can make the estimated failure probability an average over changing conditions rather than a stable property of the test. Dependence also invalidates the IID formula for the probability of at least one failure, so the resulting rerun budgets may be optimistic or otherwise inaccurate depending on the dependence structure. We therefore treat the analysis in \Cref{sec:theoretical_true_p} as an IID baseline rather than a guaranteed environment-independent budget.

Second, unlike \citet{lam2019idflakies,bell2018deflaker}, we do not reorder tests, perturb timing, vary resource stress deliberately, or inject failures. As a result, we may under-detect order-dependent, timing-sensitive, stress-induced, and environment-specific flaky tests. The reported counts and rates are therefore conservative lower bounds on behavior that a broader perturbation protocol might expose, not estimates of all flakiness in \terra.

Third, although we pin Python (3.8) and specific versions of the Qiskit and Rust toolchains, some transitive dependencies are resolved at runtime. Changes in mirror or point-release updates can introduce subtle environment differences over time. To mitigate this, we containerize each version and fix the test orchestrator configuration. However, full immutability is difficult to guarantee without fully offline and locked dependencies.

Fourth, the root-cause and fix-pattern labels rely on manual interpretation of test code, failure traces, implementation changes, and version-control history. This creates a risk of mislabeling, especially when the repository history does not explicitly discuss flakiness or when several changes occur between the last flaky release and the first stable release. To mitigate this risk, two authors independently labeled each test using the same evidence sources and reconciled disagreements through joint re-examination. Nevertheless, reconciliation does not establish category completeness, classification stability, or reproducibility by another team. Repository history may identify a change associated with the disappearance of a failure without proving the mechanism. The labels, particularly within the large \emph{Others} bucket and broad fix categories, should therefore be interpreted as exploratory best-effort classifications rather than definitive causal diagnoses.

\subsection{External}
Our results are based on the \terra component running on Linux within Singularity containers on Intel-based HPC nodes. Hardware characteristics (e.g., ISA, cache hierarchy), operating system behavior, container runtime, and CI topology all influence timing and scheduling. Other environments (e.g., macOS/Windows, AMD CPUs, cloud-based CI) may exhibit different flakiness patterns.

We restricted the release range to v.0.25.0--v.1.2.4 because of toolchain constraints. Earlier or later versions, or other quantum-computing frameworks, may behave differently~\citep{zhang2023identifying}. \terra's compiler, synthesis, and simulator-facing layers may share some concerns with comparable frameworks, but programming models, implementation languages, target gate sets, and hardware topologies can change the algorithms and failure modes. Hardware-backed runtimes and cloud-service clients may also exhibit network, queueing, calibration, service-availability, and device-related flakiness absent from this study. Generalization beyond Qiskit, beyond Python, or to hardware backends should therefore be approached with caution.

As in many software-engineering studies, the variability of real-world environments limits generalizability~\citep{wieringa2015six}. The experimental protocol is reproducible and can be applied to other frameworks, but portability of the protocol does not establish transferability of the measured rates or category proportions. We encourage the community to replicate and extend our work.

\section{Conclusions and Future Work}\label{sec:conclusions}

This paper presents a large-scale dynamic characterization of flaky tests through a longitudinal case study of Qiskit \terra. By reexecuting the \terra test suite 10,000 times across 23 releases, we provide execution-level evidence about observed prevalence, recurrence, failure intensity, detectability, and subcomponent concentration under a controlled protocol. The contribution is empirical and infrastructural rather than a new detection algorithm or statistical method.

Under the studied protocol, observed flakiness in the selected \terra releases is rare at the aggregate level but difficult to manage in practice. Many observed flaky tests have empirical failure probabilities of $10^{-3}$ or below, making them unlikely to be detected under small CI rerun budgets, while many also recur across multiple releases. These results suggest that flaky-test management in quantum-software stacks of \terra's scale should be probabilistic, historical, and budget-aware rather than relying only on release-local observations or small rerun counts. They do not establish that these behavioral patterns are unique to quantum software.

We also release a publicly available dataset containing per-test execution results for all 23 releases, providing an execution-level benchmark under the specified protocol for future work in quantum-software testing, debugging, replication, and machine-learning prediction.

Future work includes
\begin{enumerate*}[label=(\roman*)]
    \item replicating the analysis in other quantum-software frameworks and hardware-backed runtimes,
    \item using matched classical and quantum systems or controlled interventions to distinguish domain-specific mechanisms from general software mechanisms,
    \item applying timing, order, stress, and environment perturbations to expose classes of flakiness outside the present protocol,
    \item exploring more efficient or adaptive detection strategies, and
    \item developing machine-learning models that leverage both static and dynamic signals for more robust flaky-test detection and prediction.
\end{enumerate*}

\section*{Acknowledgement}
This work was partially supported by the Natural Sciences and Engineering Research Council of Canada (grant \# RGPIN-2022-03886) and the Strategic Awards for Research Transitions (START) award from University of Maryland, Baltimore County. The authors thank the Digital Research Alliance of Canada for providing computational resources. We are also grateful to Dr. Jake Lishman of the Qiskit development team for reviewing our experimental setup and preliminary findings, and for helping us validate and refine the study protocol. We are deeply grateful to the reviewers and the editors for their insightful feedback and constructive suggestions, which have substantially improved the quality and clarity of this work.

\bibliographystyle{elsarticle-harv}
\bibliography{JSS}

\begin{thebibliography}{52}
\expandafter\ifx\csname natexlab\endcsname\relax\def\natexlab#1{#1}\fi
\providecommand{\url}[1]{\texttt{#1}}
\providecommand{\href}[2]{#2}
\providecommand{\path}[1]{#1}
\providecommand{\DOIprefix}{doi:}
\providecommand{\ArXivprefix}{arXiv:}
\providecommand{\URLprefix}{URL: }
\providecommand{\Pubmedprefix}{pmid:}
\providecommand{\doi}[1]{\href{http://dx.doi.org/#1}{\path{#1}}}
\providecommand{\Pubmed}[1]{\href{pmid:#1}{\path{#1}}}
\providecommand{\bibinfo}[2]{#2}
\ifx\xfnm\relax \def\xfnm[#1]{\unskip,\space#1}\fi
\bibitem[{Alshammari et~al.(2021)Alshammari, Morris, Hilton and Bell}]{alshammari2021flakeflagger}
\bibinfo{author}{Alshammari, A.}, \bibinfo{author}{Morris, C.}, \bibinfo{author}{Hilton, M.}, \bibinfo{author}{Bell, J.}, \bibinfo{year}{2021}.
\newblock \bibinfo{title}{Flakeflagger: Predicting flakiness without rerunning tests}, in: \bibinfo{booktitle}{Proceedings of the 43rd International Conference on Software Engineering (ICSE 2021) -- Companion Volume}, \bibinfo{organization}{IEEE}. p. \bibinfo{pages}{187}.
\newblock \DOIprefix\doi{10.1109/ICSE-COMPANION52605.2021.00081}.
\bibitem[{{Anthropic}(2025)}]{anthropic2025claude}
\bibinfo{author}{{Anthropic}}, \bibinfo{year}{2025}.
\newblock \bibinfo{title}{Claude opus 4.7}.
\newblock \bibinfo{howpublished}{\url{https://www.anthropic.com/claude}}.
\newblock \bibinfo{note}{Large language model}.
\bibitem[{Bell et~al.(2018)Bell, Legunsen, Hilton, Eloussi, Yung and Marinov}]{bell2018deflaker}
\bibinfo{author}{Bell, J.}, \bibinfo{author}{Legunsen, O.}, \bibinfo{author}{Hilton, M.}, \bibinfo{author}{Eloussi, L.}, \bibinfo{author}{Yung, T.}, \bibinfo{author}{Marinov, D.}, \bibinfo{year}{2018}.
\newblock \bibinfo{title}{Deflaker: Automatically detecting flaky tests}, in: \bibinfo{booktitle}{Proceedings of the 40th International Conference on Software Engineering (ICSE '18)}, \bibinfo{organization}{ACM}. pp. \bibinfo{pages}{433--444}.
\newblock \DOIprefix\doi{10.1145/3180155.3180164}.
\bibitem[{Brown et~al.(2001)Brown, Cai and DasGupta}]{brown2001interval}
\bibinfo{author}{Brown, L.D.}, \bibinfo{author}{Cai, T.T.}, \bibinfo{author}{DasGupta, A.}, \bibinfo{year}{2001}.
\newblock \bibinfo{title}{Interval estimation for a binomial proportion}.
\newblock \bibinfo{journal}{Statistical science} \bibinfo{volume}{16}, \bibinfo{pages}{101--133}.
\bibitem[{Dutta et~al.(2020)Dutta, Shi, Choudhary, Zhang, Jain and Misailovic}]{dutta2020detecting}
\bibinfo{author}{Dutta, S.}, \bibinfo{author}{Shi, A.}, \bibinfo{author}{Choudhary, R.}, \bibinfo{author}{Zhang, Z.}, \bibinfo{author}{Jain, A.}, \bibinfo{author}{Misailovic, S.}, \bibinfo{year}{2020}.
\newblock \bibinfo{title}{Detecting flaky tests in probabilistic and machine learning applications}, in: \bibinfo{booktitle}{Proceedings of the 29th ACM SIGSOFT international symposium on software testing and analysis}, pp. \bibinfo{pages}{211--224}.
\bibitem[{Eck et~al.(2019)Eck, Palomba, Castelluccio and Bacchelli}]{eck2019understanding}
\bibinfo{author}{Eck, M.}, \bibinfo{author}{Palomba, F.}, \bibinfo{author}{Castelluccio, M.}, \bibinfo{author}{Bacchelli, A.}, \bibinfo{year}{2019}.
\newblock \bibinfo{title}{Understanding flaky tests: The developer's perspective}, in: \bibinfo{booktitle}{Proceedings of the 27th ACM Joint European Software Engineering Conference and Symposium on the Foundations of Software Engineering (ESEC/FSE '19)}, \bibinfo{organization}{ACM}. pp. \bibinfo{pages}{830--840}.
\bibitem[{Gruber and Fraser(2022)}]{gruber2022survey}
\bibinfo{author}{Gruber, M.}, \bibinfo{author}{Fraser, G.}, \bibinfo{year}{2022}.
\newblock \bibinfo{title}{A survey on how test flakiness affects developers and what support they need to address it}, in: \bibinfo{booktitle}{Proceedings of 2022 IEEE Conference on Software Testing, Verification and Validation (ICST)}, \bibinfo{organization}{IEEE}. pp. \bibinfo{pages}{82--92}.
\bibitem[{Gruber et~al.(2021)Gruber, Lukasczyk, Kroi{\ss} and Fraser}]{gruber2021empirical}
\bibinfo{author}{Gruber, M.}, \bibinfo{author}{Lukasczyk, S.}, \bibinfo{author}{Kroi{\ss}, F.}, \bibinfo{author}{Fraser, G.}, \bibinfo{year}{2021}.
\newblock \bibinfo{title}{An empirical study of flaky tests in {P}ython}, in: \bibinfo{booktitle}{2021 14th IEEE Conference on Software Testing, Verification and Validation (ICST)}, \bibinfo{organization}{IEEE}. pp. \bibinfo{pages}{148--158}.
\bibitem[{Javadi-Abhari et~al.(2024)Javadi-Abhari, Treinish, Krsulich, Wood, Lishman, Gacon, Martiel, Nation, Bishop, Cross, Johnson and Gambetta}]{qiskit2024}
\bibinfo{author}{Javadi-Abhari, A.}, \bibinfo{author}{Treinish, M.}, \bibinfo{author}{Krsulich, K.}, \bibinfo{author}{Wood, C.J.}, \bibinfo{author}{Lishman, J.}, \bibinfo{author}{Gacon, J.}, \bibinfo{author}{Martiel, S.}, \bibinfo{author}{Nation, P.D.}, \bibinfo{author}{Bishop, L.S.}, \bibinfo{author}{Cross, A.W.}, \bibinfo{author}{Johnson, B.R.}, \bibinfo{author}{Gambetta, J.M.}, \bibinfo{year}{2024}.
\newblock \bibinfo{title}{Quantum computing with {Q}iskit}.
\newblock \DOIprefix\doi{10.48550/arXiv.2405.08810}, \href{http://arxiv.org/abs/2405.08810}{{\tt arXiv:2405.08810}}.
\bibitem[{Kaur et~al.(2025)Kaur, Kim, Jamshidi and Zhang}]{kaur2025identifying}
\bibinfo{author}{Kaur, K.}, \bibinfo{author}{Kim, D.}, \bibinfo{author}{Jamshidi, A.}, \bibinfo{author}{Zhang, L.}, \bibinfo{year}{2025}.
\newblock \bibinfo{title}{Identifying flaky tests in quantum code: A machine learning approach}, in: \bibinfo{booktitle}{Proceedings of the 8th Workshop on Validation, Analysis and Evolution of Software Tests (VST 2025)}.
\newblock \DOIprefix\doi{10.48550/arXiv.2502.04471}.
\bibitem[{Khan et~al.(2025)Khan, Ye, Akbar, Khan, Mougouei and Ma}]{khan2025mining}
\bibinfo{author}{Khan, A.A.}, \bibinfo{author}{Ye, B.}, \bibinfo{author}{Akbar, M.A.}, \bibinfo{author}{Khan, J.A.}, \bibinfo{author}{Mougouei, D.}, \bibinfo{author}{Ma, X.}, \bibinfo{year}{2025}.
\newblock \bibinfo{title}{Mining q\&a platforms for empirical evidence on quantum software programming}.
\newblock \bibinfo{journal}{arXiv preprint arXiv:2503.05240} .
\bibitem[{Kurtzer et~al.(2017)Kurtzer, Sochat and Bauer}]{kurtzer2017singularity}
\bibinfo{author}{Kurtzer, G.M.}, \bibinfo{author}{Sochat, V.}, \bibinfo{author}{Bauer, M.W.}, \bibinfo{year}{2017}.
\newblock \bibinfo{title}{Singularity: Scientific containers for mobility of compute}.
\newblock \bibinfo{journal}{PloS one} \bibinfo{volume}{12}, \bibinfo{pages}{e0177459}.
\bibitem[{Lam et~al.(2019a)Lam, Godefroid, Nath, Santhiar and Thummalapenta}]{lam2019root}
\bibinfo{author}{Lam, W.}, \bibinfo{author}{Godefroid, P.}, \bibinfo{author}{Nath, S.}, \bibinfo{author}{Santhiar, A.}, \bibinfo{author}{Thummalapenta, S.}, \bibinfo{year}{2019}a.
\newblock \bibinfo{title}{Root causing flaky tests in a large-scale industrial setting}, in: \bibinfo{booktitle}{Proceedings of the 28th ACM SIGSOFT International Symposium on Software Testing and Analysis}, pp. \bibinfo{pages}{101--111}.
\bibitem[{Lam et~al.(2020)Lam, Mu{\c{s}}lu, Sajnani and Thummalapenta}]{lam2020study}
\bibinfo{author}{Lam, W.}, \bibinfo{author}{Mu{\c{s}}lu, K.}, \bibinfo{author}{Sajnani, H.}, \bibinfo{author}{Thummalapenta, S.}, \bibinfo{year}{2020}.
\newblock \bibinfo{title}{A study on the lifecycle of flaky tests}, in: \bibinfo{booktitle}{Proceedings of the ACM/IEEE 42nd International Conference on Software Engineering}, pp. \bibinfo{pages}{1471--1482}.
\bibitem[{Lam et~al.(2019b)Lam, Oei, Shi, Marinov and Xie}]{lam2019idflakies}
\bibinfo{author}{Lam, W.}, \bibinfo{author}{Oei, R.}, \bibinfo{author}{Shi, A.}, \bibinfo{author}{Marinov, D.}, \bibinfo{author}{Xie, T.}, \bibinfo{year}{2019}b.
\newblock \bibinfo{title}{idflakies: A framework for detecting and partially classifying flaky tests}, in: \bibinfo{booktitle}{Proceedings of the 12th IEEE International Conference on Software Testing, Verification and Validation (ICST 2019)}, \bibinfo{organization}{IEEE}. pp. \bibinfo{pages}{312--322}.
\newblock \DOIprefix\doi{10.1109/ICST.2019.00038}.
\bibitem[{Luo et~al.(2014)Luo, Hariri, Eloussi and Marinov}]{luo2014empirical}
\bibinfo{author}{Luo, Q.}, \bibinfo{author}{Hariri, F.}, \bibinfo{author}{Eloussi, L.}, \bibinfo{author}{Marinov, D.}, \bibinfo{year}{2014}.
\newblock \bibinfo{title}{An empirical analysis of flaky tests}, in: \bibinfo{booktitle}{Proceedings of the 22nd ACM SIGSOFT Symposium on the Foundations of Software Engineering (FSE '14)}, \bibinfo{organization}{ACM}. pp. \bibinfo{pages}{643--653}.
\bibitem[{Matsakis and Klock(2014)}]{matsakis2014rust}
\bibinfo{author}{Matsakis, N.D.}, \bibinfo{author}{Klock, F.S.}, \bibinfo{year}{2014}.
\newblock \bibinfo{title}{The rust language}, in: \bibinfo{booktitle}{Proceedings of the 2014 ACM SIGAda annual conference on High integrity language technology}, pp. \bibinfo{pages}{103--104}.
\bibitem[{Memon et~al.(2017)Memon, Gao, Nguyen, Dhanda, Nickell, Siemborski and Micco}]{memon2017taming}
\bibinfo{author}{Memon, A.}, \bibinfo{author}{Gao, Z.}, \bibinfo{author}{Nguyen, B.}, \bibinfo{author}{Dhanda, S.}, \bibinfo{author}{Nickell, E.}, \bibinfo{author}{Siemborski, R.}, \bibinfo{author}{Micco, J.}, \bibinfo{year}{2017}.
\newblock \bibinfo{title}{Taming google-scale continuous testing}, in: \bibinfo{booktitle}{Proceedings of the 2017 IEEE/ACM 39th International Conference on Software Engineering: Software Engineering in Practice Track (ICSE-SEIP)}, \bibinfo{organization}{IEEE}. pp. \bibinfo{pages}{233--242}.
\bibitem[{Micco(2017)}]{micco2017state}
\bibinfo{author}{Micco, J.}, \bibinfo{year}{2017}.
\newblock \bibinfo{title}{{The state of continuous integration testing @Google}}.
\newblock \URLprefix \url{https://research.google/pubs/the-state-of-continuous-integration-testing-google/}.
\bibitem[{Murillo et~al.(2025)Murillo, Garcia-Alonso, Moguel, Barzen, Leymann, Ali, Yue, Arcaini, P\'{e}rez-Castillo, Garc\'{\i}a-Rodr\'{\i}guez~de Guzm\'{a}n, Piattini, Ruiz-Cort\'{e}s, Brogi, Zhao, Miranskyy and Wimmer}]{murillo2025quantum}
\bibinfo{author}{Murillo, J.M.}, \bibinfo{author}{Garcia-Alonso, J.}, \bibinfo{author}{Moguel, E.}, \bibinfo{author}{Barzen, J.}, \bibinfo{author}{Leymann, F.}, \bibinfo{author}{Ali, S.}, \bibinfo{author}{Yue, T.}, \bibinfo{author}{Arcaini, P.}, \bibinfo{author}{P\'{e}rez-Castillo, R.}, \bibinfo{author}{Garc\'{\i}a-Rodr\'{\i}guez~de Guzm\'{a}n, I.}, \bibinfo{author}{Piattini, M.}, \bibinfo{author}{Ruiz-Cort\'{e}s, A.}, \bibinfo{author}{Brogi, A.}, \bibinfo{author}{Zhao, J.}, \bibinfo{author}{Miranskyy, A.}, \bibinfo{author}{Wimmer, M.}, \bibinfo{year}{2025}.
\newblock \bibinfo{title}{Quantum software engineering: Roadmap and challenges ahead}.
\newblock \bibinfo{journal}{ACM Transactions on Software Engineering and Methodology} \bibinfo{volume}{34}.
\newblock \DOIprefix\doi{10.1145/3712002}.
\bibitem[{Parry et~al.(2021)Parry, Kapfhammer, Hilton and McMinn}]{parry2021survey}
\bibinfo{author}{Parry, O.}, \bibinfo{author}{Kapfhammer, G.M.}, \bibinfo{author}{Hilton, M.}, \bibinfo{author}{McMinn, P.}, \bibinfo{year}{2021}.
\newblock \bibinfo{title}{A survey of flaky tests}.
\newblock \bibinfo{journal}{ACM Transactions on Software Engineering and Methodology (TOSEM)} \bibinfo{volume}{31}, \bibinfo{pages}{1--74}.
\bibitem[{Pinto et~al.(2020)Pinto, Miranda, Dissanayake, d'Amorim, Treude and Bertolino}]{pinto2020what}
\bibinfo{author}{Pinto, G.}, \bibinfo{author}{Miranda, B.}, \bibinfo{author}{Dissanayake, S.}, \bibinfo{author}{d'Amorim, M.}, \bibinfo{author}{Treude, C.}, \bibinfo{author}{Bertolino, A.}, \bibinfo{year}{2020}.
\newblock \bibinfo{title}{What is the vocabulary of flaky tests?}, in: \bibinfo{booktitle}{Proceedings of the 17th International Conference on Mining Software Repositories (MSR '20)}, \bibinfo{organization}{ACM/IEEE}. pp. \bibinfo{pages}{492--502}.
\newblock \DOIprefix\doi{10.1145/3379597.3387482}.
\bibitem[{{Qiskit Development Community}(2025a)}]{qiskit_configure_local}
\bibinfo{author}{{Qiskit Development Community}}, \bibinfo{year}{2025}a.
\newblock \bibinfo{title}{Configure qiskit locally}.
\newblock \URLprefix \url{https://quantum.cloud.ibm.com/docs/en/guides/configure-qiskit-local}. \bibinfo{note}{[Online; accessed 2025-11-07]}.
\bibitem[{{Qiskit Development Community}(2025b)}]{qiskit_basicaer_045}
\bibinfo{author}{{Qiskit Development Community}}, \bibinfo{year}{2025}b.
\newblock \bibinfo{title}{Qiskit basicaer provider (archived 0.45) documentation}.
\newblock \URLprefix \url{https://github.com/Qiskit/documentation/blob/archived-docs/docs/api/qiskit/0.45/providers_basicaer.mdx}. \bibinfo{note}{[Online; accessed 2025-11-07]}.
\bibitem[{{Qiskit Development Community}(2025c)}]{qiskit_circuit_api}
\bibinfo{author}{{Qiskit Development Community}}, \bibinfo{year}{2025}c.
\newblock \bibinfo{title}{Qiskit circuit api documentation}.
\newblock \URLprefix \url{https://quantum.cloud.ibm.com/docs/en/api/qiskit/circuit}. \bibinfo{note}{[Online; accessed 2025-11-07]}.
\bibitem[{{Qiskit Development Community}(2025d)}]{qiskit_compiler_api}
\bibinfo{author}{{Qiskit Development Community}}, \bibinfo{year}{2025}d.
\newblock \bibinfo{title}{Qiskit compiler api documentation}.
\newblock \URLprefix \url{https://quantum.cloud.ibm.com/docs/en/api/qiskit/compiler}. \bibinfo{note}{[Online; accessed 2025-11-07]}.
\bibitem[{{Qiskit Development Community}(2025e)}]{qiskit_quantum_info_overview}
\bibinfo{author}{{Qiskit Development Community}}, \bibinfo{year}{2025}e.
\newblock \bibinfo{title}{Qiskit operators and quantum information overview}.
\newblock \URLprefix \url{https://quantum.cloud.ibm.com/docs/en/guides/operators-overview}. \bibinfo{note}{[Online; accessed 2025-11-07]}.
\bibitem[{{Qiskit Development Community}(2025f)}]{qiskit_primitives_046}
\bibinfo{author}{{Qiskit Development Community}}, \bibinfo{year}{2025}f.
\newblock \bibinfo{title}{Qiskit primitives api documentation}.
\newblock \URLprefix \url{https://quantum.cloud.ibm.com/docs/en/api/qiskit/0.46/primitives}. \bibinfo{note}{[Online; accessed 2025-11-07]}.
\bibitem[{{Qiskit Development Community}(2025g)}]{qiskit_providers_docs}
\bibinfo{author}{{Qiskit Development Community}}, \bibinfo{year}{2025}g.
\newblock \bibinfo{title}{Qiskit providers and backend interface documentation}.
\newblock \URLprefix \url{https://quantum.cloud.ibm.com/docs/}. \bibinfo{note}{[Online; accessed 2025-11-07]}.
\bibitem[{{Qiskit Development Community}(2025h)}]{qiskit_pulse_api}
\bibinfo{author}{{Qiskit Development Community}}, \bibinfo{year}{2025}h.
\newblock \bibinfo{title}{Qiskit pulse api and guide}.
\newblock \URLprefix \url{https://quantum.cloud.ibm.com/docs/en/api/qiskit/pulse}. \bibinfo{note}{[Online; accessed 2025-11-07]}.
\bibitem[{{Qiskit Development Community}(2025i)}]{qiskit_qpy_api}
\bibinfo{author}{{Qiskit Development Community}}, \bibinfo{year}{2025}i.
\newblock \bibinfo{title}{Qiskit qpy api documentation}.
\newblock \URLprefix \url{https://quantum.cloud.ibm.com/docs/en/api/qiskit/qpy}. \bibinfo{note}{[Online; accessed 2025-11-07]}.
\bibitem[{{Qiskit Development Community}(2025j)}]{qiskit_description}
\bibinfo{author}{{Qiskit Development Community}}, \bibinfo{year}{2025}j.
\newblock \bibinfo{title}{Qiskit readme documentation}.
\newblock \URLprefix \url{https://github.com/Qiskit/qiskit/blob/25c8a9312b9f92d54c05c27aa32ecd30cae15301/README.md?plain=1#L15-L16}.
\bibitem[{{Qiskit Development Community}(2025k)}]{qiskit_repo_test_examples}
\bibinfo{author}{{Qiskit Development Community}}, \bibinfo{year}{2025}k.
\newblock \bibinfo{title}{Qiskit repository and \texttt{test\_examples} test suite}.
\newblock \URLprefix \url{https://github.com/Qiskit/qiskit}. \bibinfo{note}{[Online; accessed 2025-11-07]}.
\bibitem[{{Qiskit Development Community}(2025l)}]{qiskit_scheduler_045}
\bibinfo{author}{{Qiskit Development Community}}, \bibinfo{year}{2025}l.
\newblock \bibinfo{title}{Qiskit scheduler (archived 0.45) documentation}.
\newblock \URLprefix \url{https://github.com/Qiskit/documentation/blob/archived-docs/docs/api/qiskit/0.45/scheduler.mdx}. \bibinfo{note}{[Online; accessed 2025-11-07]}.
\bibitem[{{Qiskit Development Community}(2025m)}]{qiskit_synthesis_api}
\bibinfo{author}{{Qiskit Development Community}}, \bibinfo{year}{2025}m.
\newblock \bibinfo{title}{Qiskit synthesis api documentation}.
\newblock \URLprefix \url{https://quantum.cloud.ibm.com/docs/en/api/qiskit/synthesis}. \bibinfo{note}{[Online; accessed 2025-11-07]}.
\bibitem[{{Qiskit Development Community}(2025n)}]{qiskit_opflow_045}
\bibinfo{author}{{Qiskit Development Community}}, \bibinfo{year}{2025}n.
\newblock \bibinfo{title}{Qiskit \texttt{opflow} (archived 0.45) documentation}.
\newblock \URLprefix \url{https://github.com/Qiskit/documentation/blob/archived-docs/docs/api/qiskit/0.45/opflow.mdx}. \bibinfo{note}{[Online; accessed 2025-11-07]}.
\bibitem[{{Qiskit Development Community}(2025o)}]{qiskit_quantum_info_api}
\bibinfo{author}{{Qiskit Development Community}}, \bibinfo{year}{2025}o.
\newblock \bibinfo{title}{Qiskit \texttt{quantum\_info} api documentation}.
\newblock \URLprefix \url{https://quantum.cloud.ibm.com/docs/en/api/qiskit/quantum_info}. \bibinfo{note}{[Online; accessed 2025-11-07]}.
\bibitem[{{Qiskit Development Community}(2025p)}]{qiskit_transpiler_api}
\bibinfo{author}{{Qiskit Development Community}}, \bibinfo{year}{2025}p.
\newblock \bibinfo{title}{Qiskit transpiler api documentation}.
\newblock \URLprefix \url{https://quantum.cloud.ibm.com/docs/en/api/qiskit/transpiler}. \bibinfo{note}{[Online; accessed 2025-11-07]}.
\bibitem[{{Qiskit Development Community}(2025q)}]{qiskit_visualization_api}
\bibinfo{author}{{Qiskit Development Community}}, \bibinfo{year}{2025}q.
\newblock \bibinfo{title}{Qiskit visualization api documentation}.
\newblock \URLprefix \url{https://quantum.cloud.ibm.com/docs/en/api/qiskit/visualization}. \bibinfo{note}{[Online; accessed 2025-11-07]}.
\bibitem[{{Qiskit Development Community}(2026a)}]{qiskit2026validation_discussion}
\bibinfo{author}{{Qiskit Development Community}}, \bibinfo{year}{2026}a.
\newblock \bibinfo{title}{Empirical snapshot of flaky tests in qiskit v2.2.3; feedback welcome · issue \#15605 · qiskit/qiskit}.
\newblock \URLprefix \url{https://github.com/Qiskit/qiskit/issues/15605}. \bibinfo{note}{[Online; accessed 2026-05-28]}.
\bibitem[{{Qiskit Development Community}(2026b)}]{qiskit2026fix}
\bibinfo{author}{{Qiskit Development Community}}, \bibinfo{year}{2026}b.
\newblock \bibinfo{title}{Seed randomness in `ucr' tests by jakelishman · pull request \#15651 · qiskit/qiskit}.
\newblock \URLprefix \url{https://github.com/Qiskit/qiskit/pull/15651}. \bibinfo{note}{[Online; accessed 2026-05-28]}.
\bibitem[{Sivaloganathan et~al.(2024)Sivaloganathan, Jamshidi, Miranskyy and Zhang}]{sivaloganathan2024automating}
\bibinfo{author}{Sivaloganathan, J.}, \bibinfo{author}{Jamshidi, A.}, \bibinfo{author}{Miranskyy, A.}, \bibinfo{author}{Zhang, L.}, \bibinfo{year}{2024}.
\newblock \bibinfo{title}{Automating quantum software maintenance: Flakiness detection and root cause analysis}.
\newblock \bibinfo{journal}{arXiv preprint arXiv:2410.23578} .
\bibitem[{Sivaloganathan et~al.(2026)Sivaloganathan, Jamshidi, Miranskyy and Zhang}]{sivaloganathan2026automating}
\bibinfo{author}{Sivaloganathan, J.}, \bibinfo{author}{Jamshidi, A.}, \bibinfo{author}{Miranskyy, A.}, \bibinfo{author}{Zhang, L.}, \bibinfo{year}{2026}.
\newblock \bibinfo{title}{Automating detection and root-cause analysis of flaky tests in quantum software}.
\newblock \bibinfo{journal}{arXiv preprint arXiv:2603.09029} .
\bibitem[{{Tox Development Community}()}]{tox}
\bibinfo{author}{{Tox Development Community}}, .
\newblock \bibinfo{title}{tox-dev/tox: Command line driven {CI} frontend and development task automation tool.}
\newblock \URLprefix \url{https://github.com/tox-dev/tox}.
\bibitem[{Verdecchia et~al.(2021)Verdecchia, Cruciani, Miranda and Bertolino}]{verdecchia2021know}
\bibinfo{author}{Verdecchia, R.}, \bibinfo{author}{Cruciani, E.}, \bibinfo{author}{Miranda, B.}, \bibinfo{author}{Bertolino, A.}, \bibinfo{year}{2021}.
\newblock \bibinfo{title}{Know you neighbor: Fast static prediction of test flakiness}.
\newblock \bibinfo{journal}{IEEE Access} \bibinfo{volume}{9}, \bibinfo{pages}{76119--76134}.
\bibitem[{Wieringa and Daneva(2015)}]{wieringa2015six}
\bibinfo{author}{Wieringa, R.J.}, \bibinfo{author}{Daneva, M.}, \bibinfo{year}{2015}.
\newblock \bibinfo{title}{Six strategies for generalizing software engineering theories}.
\newblock \bibinfo{journal}{Science of computer programming} \bibinfo{volume}{101}, \bibinfo{pages}{136--152}.
\newblock \DOIprefix\doi{10.1016/J.SCICO.2014.11.013}.
\bibitem[{Wilson(1927)}]{wilson1927probable}
\bibinfo{author}{Wilson, E.B.}, \bibinfo{year}{1927}.
\newblock \bibinfo{title}{Probable inference, the law of succession, and statistical inference}.
\newblock \bibinfo{journal}{Journal of the American Statistical Association} \bibinfo{volume}{22}, \bibinfo{pages}{209--212}.
\bibitem[{Wohlin et~al.(2012)Wohlin, Runeson, H{\"o}st, Ohlsson, Regnell and Wessl{\'e}n}]{wohlin2012experimentation}
\bibinfo{author}{Wohlin, C.}, \bibinfo{author}{Runeson, P.}, \bibinfo{author}{H{\"o}st, M.}, \bibinfo{author}{Ohlsson, M.}, \bibinfo{author}{Regnell, B.}, \bibinfo{author}{Wessl{\'e}n, A.}, \bibinfo{year}{2012}.
\newblock \bibinfo{title}{Experimentation in Software Engineering}.
\newblock Computer Science, \bibinfo{publisher}{Springer Berlin Heidelberg}.
\bibitem[{Yin(2009)}]{yin2009case}
\bibinfo{author}{Yin, R.}, \bibinfo{year}{2009}.
\newblock \bibinfo{title}{Case Study Research: Design and Methods}.
\newblock Applied Social Research Methods, \bibinfo{publisher}{SAGE Publications}.
\bibitem[{Zhang and Miranskyy(2024)}]{zhang2024automated}
\bibinfo{author}{Zhang, L.}, \bibinfo{author}{Miranskyy, A.}, \bibinfo{year}{2024}.
\newblock \bibinfo{title}{Automated flakiness detection in quantum software bug reports}, in: \bibinfo{booktitle}{2024 IEEE International Conference on Quantum Computing and Engineering (QCE)}, \bibinfo{organization}{IEEE}. pp. \bibinfo{pages}{179--181}.
\bibitem[{Zhang et~al.(2023)Zhang, Radnejad and Miranskyy}]{zhang2023identifying}
\bibinfo{author}{Zhang, L.}, \bibinfo{author}{Radnejad, M.}, \bibinfo{author}{Miranskyy, A.}, \bibinfo{year}{2023}.
\newblock \bibinfo{title}{Identifying flakiness in quantum programs}, in: \bibinfo{booktitle}{Proceedings of the 17th ACM/IEEE International Symposium on Empirical Software Engineering and Measurement (ESEM 2023)}, \bibinfo{organization}{ACM/IEEE}.
\newblock \DOIprefix\doi{10.1109/ESEM56168.2023.10304850}.
\bibitem[{Ziftci and Cavalcanti(2020)}]{ziftci2020flake}
\bibinfo{author}{Ziftci, C.}, \bibinfo{author}{Cavalcanti, D.}, \bibinfo{year}{2020}.
\newblock \bibinfo{title}{De-flake your tests: Automatically locating root causes of flaky tests in code at {G}oogle}, in: \bibinfo{booktitle}{2020 IEEE International Conference on Software Maintenance and Evolution (ICSME)}, \bibinfo{organization}{IEEE}. pp. \bibinfo{pages}{736--745}.

\end{thebibliography}
\clearpage
\appendix

\section{Module taxonomy}\label{app:modules}
Below are descriptions of the main subcomponents of Qiskit \terra.

\begin{description}
  \item[\texttt{transpiler}]
  Tests passes, preset pass managers, layout, routing, basis translation, scheduling, and \texttt{transpile()} behavior.
  This subcomponent is present and actively maintained across all studied releases (0.25.0--1.2.4)~\citep{qiskit_transpiler_api}.

  \item[\texttt{quantum\_info}]
  Tests the state/operator/channel toolbox (e.g., \texttt{Statevector}, \texttt{DensityMatrix}, \texttt{SparsePauliOp}), including conversions, algebraic operations, metrics, and channel properties.
  Available throughout 0.25.0--1.2.4~\citep{qiskit_quantum_info_overview,qiskit_quantum_info_api}.

  \item[\texttt{compiler}]
  Tests high-level compile wrappers (e.g., \texttt{transpile}, \texttt{assemble}) that orchestrate transpilation, scheduling, and backend configuration.
  Present as a thin wrapper interface in all considered versions~\citep{qiskit_compiler_api}.

  \item[\texttt{synthesis}]
  Tests unitary and operator synthesis routines that decompose targets into supported gate sets (exact and approximate).
  Supported across 0.25.0--1.2.4~\citep{qiskit_synthesis_api}.

  \item[\texttt{circuit}]
  Tests the core \texttt{QuantumCircuit} abstraction, gate and instruction definitions, parameterization, control-flow, and circuit transformation APIs.
  Central and stable over 0.25.0--1.2.4~\citep{qiskit_circuit_api}.

  \item[\texttt{qpy}]
  Tests QPY-based serialization and deserialization of circuits, checking round-trip correctness and compatibility.
  Used across all studied releases~\citep{qiskit_qpy_api}.

  \item[\texttt{pulse}]
  Tests pulse-level programming (channels, waveforms, schedule blocks) and alignment with backend timing/constraints.
  Available throughout 0.25.0--1.2.4, though some features are later streamlined~\citep{qiskit_pulse_api}.

  \item[\texttt{scheduler}]
  Tests legacy helpers converting circuits to scheduled pulse programs, including timing and resource constraints.
  Relevant only to earlier releases in our range (removed around 0.45)~\citep{qiskit_scheduler_045}.

  \item[\texttt{visualization}]
  Tests circuit drawers, state plots, device/heatmap plots, and related options.
  Present from 0.25.0 to 1.2.4~\citep{qiskit_visualization_api}.

  \item[\texttt{providers}]
  Tests provider and backend interfaces (including fake backends), job submission, configuration, and result formats.
  Present across all studied versions~\citep{qiskit_providers_docs}.

  \item[\texttt{test\_user\_config}]
  Tests loading and applying user configuration (e.g., defaults, visualization settings) from local config files.
  Present in our full range~\citep{qiskit_configure_local}.

  \item[\texttt{basicaer}]
  Tests legacy local simulators for correctness and compatibility.
  Only applicable to the subset 0.25.0--0.46.x within our window~\citep{qiskit_basicaer_045}.

  \item[\texttt{opflow}]
  Tests operator-flow abstractions for operator expressions, expectations, and gradients (now legacy).
  Appears in 0.25.0--0.46.x within our range~\citep{qiskit_opflow_045}.

  \item[\texttt{primitives}]
  Tests the \texttt{Sampler} and \texttt{Estimator} primitives, including interfaces, result formats, and backend/runtime integration.
  Becomes relevant from roughly 0.39.0 and remains central through 1.2.4~\citep{qiskit_primitives_046}.

  \item[\texttt{test\_examples}]
  Tests example scripts and tutorials to ensure that published examples run end-to-end against the public API.
  Present throughout 0.25.0--1.2.4~\citep{qiskit_repo_test_examples}.
\end{description}

\end{document}